\documentclass[a4paper,10pt]{article}
\usepackage{amssymb}
\usepackage{latexsym}
\usepackage{amsmath}
\usepackage{amscd}

\numberwithin{equation}{section}

\newcommand{\bR}{{\mathbb R}}

\newcommand{\bC}{{\mathbb C}}

\newcommand{\kC}{{\mathcal C}}
\newcommand{\kD}{{\mathcal D}}

\newcommand{\kH}{{\mathcal H}}

\newcommand{\kN}{{\mathcal N}}

\newcommand{\gotG}{{\mathfrak G}}

\newcommand{\gotH}{{\mathfrak H}}

\newcommand{\gotK}{{\mathfrak K}}

\newcommand{\gotL}{{\mathfrak L}}

\newcommand{\gotS}{{\mathfrak S}}

\newcommand{\gga}{{\gamma}}
\newcommand{\gG}{{\Gamma}}

\newcommand{\gl}{{\lambda}}

\newcommand{\gO}{{\Omega}}
\newcommand{\gP}{{\Pi}}

\newcommand{\gt}{{\tau}}

\newcommand{\gT}{{\Theta}}

\newcommand{\gY}{{\Upsilon}}

\newcommand{\slim}{\,\mbox{\rm s-}\hspace{-2pt} \lim}

\newcommand{\real}{{\Re{\mathrm e\,}}}
\newcommand{\imag}{{\Im{\mathrm m\,}}}
\newcommand{\dom}{{\mathrm{dom\,}}}
\newcommand{\ran}{{\mathrm{ran\,}}}

\newcommand{\tr}{{\mathrm{tr}}}

\newcommand{\clo}{{\mathrm{clo}}}

\newcommand{\spa}{{\mathrm{span}}}
\newcommand{\clospa}{{\mathrm{clospan}}}

\newcommand{\wt}[1]{{\widetilde{#1}}}

\newtheorem{thm}{Theorem}[section]
\newtheorem{prop}[thm]{Proposition}
\newtheorem{lem}[thm]{Lemma}
\newtheorem{cor}[thm]{Corollary}

\newtheorem{defn}[thm]{Definition}

\newcommand{\ba}{\begin{array}}
\newcommand{\ea}{\end{array}}
\newcommand{\bea}{\begin{eqnarray}}
\newcommand{\eea}{\end{eqnarray}}
\newcommand{\bead}{\begin{eqnarray*}}
\newcommand{\eead}{\end{eqnarray*}}
\newcommand{\be}{\begin{equation}}
\newcommand{\ee}{\end{equation}}
\newcommand{\bed}{\begin{displaymath}}
\newcommand{\eed}{\end{displaymath}}
\newcommand{\bl}{\begin{lem}}
\newcommand{\el}{\end{lem}}
\newcommand{\bp}{\begin{prop}}
\newcommand{\ep}{\end{prop}}
\newcommand{\bt}{\begin{thm}}
\newcommand{\et}{\end{thm}}
\newcommand{\Label}{\label}
\newcommand{\bc}{\begin{cor}}
\newcommand{\ec}{\end{cor}}
\newcommand{\la}{\Label}

\newcommand{\bd}{\begin{defn}}
\newcommand{\ed}{\end{defn}}

\newenvironment{proof}%
{\begin{sloppypar}\noindent{\bf Proof.}}%
{\hspace*{\fill}$\square$\end{sloppypar}\bigskip}

\def\sG{{\mathfrak G}}   \def\sH{{\mathfrak H}}   
   \def\sK{{\mathfrak K}}   \def\sL{{\mathfrak L}}

\def\sS{{\mathfrak S}}

      \def\dC{{\mathbb C}}

      \def\dR{{\mathbb R}}

   \def\dZ{{\mathbb Z}}

      \def\cC{{\mathcal C}}
\def\cD{{\mathcal D}}      
\def\cG{{\mathcal G}}   \def\cH{{\mathcal H}}   
   \def\cK{{\mathcal K}}

\def\sp{{\text{\rm span\,}}}

\def\op{{\text{\rm op\,}}}

\title{Trace formulae for dissipative and coupled\\ scattering
  systems}

\author{Jussi Behrndt\thanks{E-mail: behrndt@math.tu-berlin.de}\\
Technische Universit\"{a}t Berlin\\
Institut f\"{u}r Mathematik MA 6--4\\
Stra\ss e des 17.\ Juni 136\\
 D--10623 Berlin\\
Germany
\and
Mark M. Malamud\thanks{mmm@telenet.dn.ua}\\
Donetsk National University\\
Department of Mathematics\\
Universitetskaya 24\\
83055 Donetsk\\
Ukraine
\and
Hagen Neidhardt\thanks{neidhardt@wias-berlin.de}\\
Weierstra{\ss}-Institut f\"ur\\
Angewandte Analysis und Stochastik\\
Mohrenstr. 39\\
D-10117 Berlin\\
Germany
} 

\begin{document}

\maketitle

\begin{abstract}
For scattering systems consisting of a 
(family of) maximal dissipative extension(s) 
and a selfadjoint extension of a symmetric operator with finite 
deficiency indices, 
the spectral shift function is expressed in terms of an abstract 
Titchmarsh-Weyl function and a variant of the Birman-Krein formula is proved.
\end{abstract}

{\bf 2000 Mathematics Subject Classification:} 47A40, 47A55, 47B44\\

{\bf Keywords:}
Scattering system, scattering matrix, boundary triplet,
Titchmarsh-Weyl function, spectral shift function,
Krein-Birman formula

\date{December 19, 2007}

\maketitle

\tableofcontents

\section{Introduction}

The main objective of this paper is to apply and to extend results from 
\cite{BeMN1}  and \cite{BeMN2} on scattering 
matrices and spectral shift functions for 
pairs of selfadjoint or maximal dissipative extensions of a symmetric operator $A$ with finite deficiency indices 
in a Hilbert space $\mathfrak H$.

Let us first briefly recall some basic concepts.
For a pair of selfadjoint operators $H$ and $H_0$
in $\mathfrak H$ the wave operators
$W_{\pm}(H,H_0)$ of the scattering system $\{H,H_0\}$  are defined by
\bed
W_{\pm}(H,H_0) = s-\lim_{t\to\pm\infty} e^{iHt}e^{-iH_0t}P_{ac}(H_0),
\eed
where $P^{ac}(H_0)$ is the projection onto the absolutely continuous
subspace of the unperturbed operator $H_0$. If for instance the resolvent difference
\begin{equation}\label{resdifftrace}
(H-z)^{-1} - (H_0-z)^{-1}\in\sS_1,\qquad  z\in \rho(H)\cap\rho(H_0)
\end{equation}
is a trace class operator, then it is well known that the wave operators $W_{\pm}(H,H_0)$ 
exist and are isometric in $\sH$, see, e.g. \cite{Y}.
The scattering operator $S(H,H_0)$ of the scattering system $\{H,H_0\}$ is defined by
\bed
S(H,H_0) = W_{+}(H,H_0)^*W_{-}(H,H_0).
\eed
$S(H,H_0)$ commutes with $H_0$ and is unitary on the absolutely continuous
subspace of $H_0$. Therefore
$S(H,H_0)$ is unitarily equivalent to a multiplication operator induced by a family  
$S(H,H_0; \lambda)$ of unitary operators in the spectral
representation of $H_0$. This family is usually called the scattering
matrix of the scattering system $\{H,H_0\}$ and 
is one of the most important quantities in the analysis of scattering
processes. 
%Since the spectral representation of $H_0$ is unique
%only up to pointwise unitary equivalence the scattering matrix
%is unique up to pointwise unitary equivalence.

Another important object in scattering and perturbation theory is the so-called
spectral shift function introduced by M.G. Krein in
\cite{K62}. For the case $\dom(H)  = \dom(H_0)$  and 
$V = H-H_0\in {\mathfrak S}_1$ a spectral shift function $\xi$ of the pair $\{H,H_0\}$ 
was defined with the help of the perturbation determinant 
\begin{equation}\label{0.4}
D_{H/H_0}(z) := \det\left((H-z)(H_0-z)^{-1}\right).
\end{equation}
Since $\lim_{|\imag(z)|\to \infty} D_{H/H_0}(z)=1 $ 
a branch of $z\mapsto \ln(D_{H/H_0}(z))$ in the upper half plane ${\mathbb C}_+$ 
is fixed by the condition  $\ln(D_{H/H_0}(z))\to 0$ as $\imag(z) \to \infty$ and
the  spectral shift function is then defined by
\begin{equation}\label{0.3}
\xi(\lambda) = \frac{1}{\pi}\imag\bigl(\ln\left(D_{H/H_0}(\lambda + i0)\right)\bigr)
  = \frac{1}{\pi}\lim_{\varepsilon\to 0}\imag\bigl(\ln\left(D_{H/H_0}(\lambda + i\varepsilon)\right)\bigr).
\end{equation}
M.G.~Krein proved that $\xi\in L_1(\bR,d\gl)$,  $\|\xi\|_{L_1}\le \|V\|_1$, and that the trace
formula
\begin{equation}\label{0.5}
\tr\left((H-z)^{-1}- (H_0-z)^{-1}\right) = - \int_{\mathbb R}\frac{\xi(\lambda)}{(\lambda -z)^{2}}\, d\lambda
\end{equation}
holds for all $z\in\rho(H)\cap\rho(H_0)$. It turns out that the scattering matrix 
and the spectral shift function of the pair $\{H,H_0\}$ are related via the
Birman-Krein formula:
\begin{equation}\label{0.6}
\det\bigl(S(H,H_0; \lambda)\bigr) =  \exp \bigl(-2\pi i\xi(\lambda)\bigr) \qquad 
\text{for a.e.} \quad \lambda \in \mathbb R.
\end{equation}
The trace formula and the Birman-Krein formula can be extended to the case
that only the resolvent difference \eqref{resdifftrace} of $H$ and $H_0$ is trace class. Namely, then
there exists a real measurable function $\xi\in L^1(\bR,(1+\lambda^2)^{-1}d\gl)$
such that \eqref{0.5} and \eqref{0.6} hold. However, in this situation it is not immediately clear 
how the perturbation determinant in \eqref{0.3} has to be replaced. 

%Observe 
%that for such pairs the assumption  $\dom(H) = \dom(H_0)$ is, in general, not
%satisfied. 
%Then there is a real measurable function
%$\xi_{H/H_0}(\cdot) \in L^1(\bR,(1+\lambda^2)^{-1}d\gl)$
%such that the trace formula \eqref{0.5} holds. The usual way to prove
%this is to carry over the problem to unitary operators using 
%the Cayley transforms $U=(H-i)(H+i)^{-1}$ and $U_0 = (H_0 -i)(H_0 + i)^{-1}$ of  $H$ and
%$H_0$ respectively (see \cite{K62,Y}) and to prove a trace formula for
%unitary operators. 
%We note that the spectral
%shift function of $\{H,H_0\}$ is determined by the trace formula \eqref{0.5} up to a real constant. Choosing an
%appropriate real constant it is possible to prove the Birman-Krein
%formula \eqref{0.6}. 
%Trace formula and Birman-Krein formula fix the spectral
%shift function up to an integer. However, the problem remains open whether
%it is possible to find an analogon for the perturbation determinant.

In Section~\ref{two} we propose a possible solution of this problem for pairs
of selfadjoint extensions $A_0$ and $A_\Theta$ of a densely defined
symmetric operator $A$ with finite deficiency indices. 
Observe that here the resolvent difference is even a finite rank operator.
In order to describe the pair $\{A_\Theta,A_0\}$ and a corresponding spectral shift function we use the notion
of boundary triplets and associated Weyl functions. More precisely, we choose a boundary triplet
$\Pi=\{\cH, \Gamma_0, \Gamma_1\}$ for $A^*$ and a selfadjoint parameter $\Theta$ in $\cH$ such that
$A_0 = A^*\upharpoonright\ker(\gG_0)$ and $A_{\Theta} = A^*\upharpoonright\ker(\Gamma_1- \Theta\Gamma_0)$ holds.
If $M(\cdot)$ is the Weyl function associated with this boundary triplet
it is shown in Theorem~\ref{V.1} (see also \cite{BeMN1} and \cite{LSY} for special cases)
%that the Hilbert space
%$L^2(\bR,\kH_\gl,d\gl)$, $\kH_\gl := \ran(\imag(M(\gl+i0))$, performs a
%spectral representation of the absolutely part $A^{ac}_0$ of $A_0$ and
%that the scattering matrix $S_\gT(\cdot) := S(A_\gT, A_0; \cdot)$ of the scattering system $\{A_\gT,A_0\}$
%admits the representation 
%
%
%\begin{equation}\label{0.7}
%S_\gT(\gl) =
%I_{\kH_{M(\gl)}} +
%2i\sqrt{\imag(M(\gl))}\bigl(\gT-M(\gl)\bigr)^{-1}
%\sqrt{\imag(M(\gl))},
%\end{equation}
%
%
%see also \cite{AP1} by V.M.~Adamyan and B.S.~Pavlov where a similar type of representation 
%was found.
%Moreover, it was proved in \cite{BeMN1} 
that a spectral shift
function $\xi(\cdot)$ of the pair $\{A_\Theta,A_0\}$ can be chosen as
\begin{equation}\label{0.7a}
 \begin{split}
  \xi_\gT(\lambda) & =\frac{1}{\pi}\imag\bigl(\tr\bigl(\log(M(\gl + i0) - \gT)\bigr)\bigr)\\
&= \frac{1}{\pi}\imag\bigl(\ln\bigl(\det(M(\gl + i0) - \gT)\bigr)\bigr)
+ 2k,\quad k\in\dZ.
 \end{split}
\end{equation}
%
%
%$k \in \bZ$, for an arbitrary branch of $\ln(\cdot)$.
By comparing \eqref{0.7a} with \eqref{0.3} it is clear that 
$\det\bigl(M(z) - \gT\bigr)$  plays a similar role as
the perturbation determinant \eqref{0.4} for additive perturbations.
Moreover, a simple proof of the Birman-Krein formula \eqref{0.6} in this situation
is obtained in Section~\ref{ssfunction} by using the representation 
\begin{equation}\label{0.7}
S_\gT(\gl) =
I_{\kH_{M(\gl)}} +
2i\sqrt{\imag(M(\gl))}\bigl(\gT-M(\gl)\bigr)^{-1}
\sqrt{\imag(M(\gl))}
\end{equation}
of the scattering matrix $S_\gT(\cdot) = S(A_\gT, A_0; \cdot)$ 
of the scattering system $\{A_\gT,A_0\}$
from \cite{BeMN1}, cf. also the work \cite{AP1} by V.M.~Adamyan and B.S.~Pavlov.
%it was proved in \cite{BeMN1} 
%that the Hilbert space
%$L^2(\bR,\kH_\gl,d\gl)$, $\kH_\gl := \ran(\imag(M(\gl+i0))$, performs a
%spectral representation of the absolutely part $A^{ac}_0$ of $A_0$ and
%that the scattering matrix $S_\gT(\cdot) := S(A_\gT, A_0; \cdot)$ 
%of the scattering system $\{A_\gT,A_0\}$
%admits the representation 
%
%
%\begin{equation}\label{0.7}
%S_\gT(\gl) =
%I_{\kH_{M(\gl)}} +
%2i\sqrt{\imag(M(\gl))}\bigl(\gT-M(\gl)\bigr)^{-1}
%\sqrt{\imag(M(\gl))},
%\end{equation}
%
%
%cf. also \cite{AP1} by V.M.~Adamyan and B.S.~Pavlov.
%Combining \eqref{0.7} and \eqref{0.7a} one gets a simple proof of the Birman-Krein formula.

These results are generalized to maximal dissipative extensions in Section~\ref{dilations}. 
Let again $A$ be a symmetric operator in $\gotH$ with finite
deficiency and let 
$\Pi=\{\cH, \Gamma_0, \Gamma_1\}$  be a boundary triplet for $A^*$. 
If $D$ is a dissipative matrix in $\cH$, $\imag(D) \le 0$, then 
$A_{D} = A^*\upharpoonright\ker(\Gamma_1- D\Gamma_0)$ is a maximal
dissipative extension of $A$. For the scattering system
$\{A_D,A_0\}$ the wave operators  $W_{\pm}(A_D, A_0)$, the scattering operator $S(A_D, A_0)$
and the scattering matrix $S(A_D, A_0; \gl)$ can be defined
similarly as in the selfadjoint case. It turns out that the representation \eqref{0.7} extends to the dissipative
case. More precisely, the Hilbert space
$L^2(\bR,\kH_\gl,d\gl)$, $\kH_\gl := \ran(\imag(M(\gl+i0))$, performs a
spectral representation of the absolutely part $A^{ac}_0$ of $A_0$ and
the scattering matrix $S_D(\cdot) := S(A_D, A_0; \cdot)$ of the scattering system $\{A_D,A_0\}$
admits the representation 
\begin{equation*}
S_D(\gl) =
I_{\kH_{M(\gl)}} +
2i\sqrt{\imag(M(\gl))}\bigl(D - M(\gl)\bigr)^{-1}
\sqrt{\imag(M(\gl))},
\end{equation*}
cf. \cite[Theorem 3.8]{BeMN2}. With the help of a minimal selfadjoint dilation $\widetilde K$ 
of $A_D$ in the Hilbert space $\sH\oplus L^2(\dR,\cH_D)$, $\cH_D:=\ran(\imag(D))$, we verify that there is 
a spectral shift function $\eta_D$
of the pair $\{A_D,A_0\}$ such that 
the trace formula
\bed
\tr\left((A_D - z)^{-1} - (A_0 - z)^{-1}\right) = -\int_\bR \frac{\eta_D(\gl)}{(\gl - z)^{2}}\,d\gl,\quad z\in\dC_+,
\eed
%
%
%holds wit $z$ is restricted to the upper have plane $\bC_+$. 
holds and this spectral
shift function $\eta_D(\cdot)$ admits the representation
\bead
\eta_D(\lambda) & = &
\frac{1}{\pi}\imag\bigl(\tr\bigl(\log(M(\gl + i0) - D)\bigr)\bigr)\\
& =  & \frac{1}{\pi}\imag\bigl(\ln\bigl(\det(M(\gl + i0) - D)\bigr)\bigr)
+ 2k,\quad k\in\dZ,\nonumber
\eead
cf. Theorem~\ref{III.8}. In Section~\ref{modbk} we show that the Birman-Krein formula holds in the
modified form
\bed
\det(S_D(\gl)) = \det(W_{A_D}(\gl-i0)))\exp\bigl(-2\pi i \eta_D(\gl)\bigr)
\eed
for a.e. $\gl \in \bR$, where $z\mapsto W_{A_D}(z)$, $z \in \bC_-$, is the
characteristic function of the maximal dissipative operator
$A_D$. Since by \cite{AA1,AA2,AA3,AA4} the limit
$W_{A_D}(\gl-i0)^*$ can be regarded as the scattering matrix
$S^{LP}(\cdot)$ of an appropriate Lax-Phillips scattering system one
gets finally the representation
\begin{equation}\label{0.8}
\det(S_D(\gl)) = \overline{\det(S^{LP}(\gl)))}\exp\bigl(-2\pi i \eta_D(\gl)\bigr)
\end{equation}
for a.e. $\gl \in \bR$. The results correspond to similar results for
additive dissipative perturbations,
\cite{AN90,N86,N87,N88,Ryb83,Ryb84,Ryb85}.

In Section~\ref{IV} so-called open quantum system with finite rank coupling are investigated.
Here we follow the lines of \cite{BeMN2}. From the mathematical point of view these open quantum 
systems are closely related to the Krein-Naimark formula for 
generalized resolvents and the \v{S}trauss family of extensions of a symmetric operator.
Recall that the Krein-Naimark formula establishes a one-to-one correspondence
between the generalized resolvents
$z\mapsto P_{\mathfrak H}(\widetilde L - z)^{-1}\upharpoonright\gotH$ of the symmetric operator $A$,
that is, the compressed resolvents of selfadjoint extensions $\widetilde L$ of $A$ in bigger Hilbert spaces, 
and the class of Nevanlinna families $\tau(\cdot)$ via
\bed
 P_{\mathfrak H}(\wt L - z)^{-1}\upharpoonright\gotH = 
(A_0-z)^{-1} - \gamma(z) (\tau(z) + M(z))^{-1}\gamma(\bar z)^*.
\eed
Here $\Pi_A = \{\cH, \Gamma_0, \Gamma_1\}$ is a boundary triplet for $A^*$ and $\gamma$ and $M$ are the
corresponding $\gamma$-field and Weyl function, respectively. 
%We note that Nevanlinna function
%$\gt(\cdot)$ can be obtained from the Naimark-Krein formula  
%$-\tau(z) = \Gamma P_{\mathfrak H}(\wt L - z)^{-1}\upharpoonright\gotH$ 
%where $\Gamma :=\{\Gamma_0, \Gamma_1\}$. 
%In general, $\tau(z)$ is a family of (maximal dissipative) relations in $\cH$.
%If $\tau(z)$ is a family of (maximal dissipative) operators in $\cH$, then the latter relation
%can be rewritten as
%
%
It can be shown that the generalized resolvent coincides pointwise with the resolvent of the \v{S}trauss extension 
\begin{equation*}
A_{-\gt(z)} := A^*\upharpoonright\ker\bigl(\gG_1 + \gt(z)\gG_0\bigr),
\end{equation*}
i.e., $P_{\mathfrak H}(\wt L - z)^{-1}\upharpoonright\gotH = (A_{-\gt(z)} - z)^{-1}$ holds, and that for 
$z\in\dC_+$ each extension  $A_{-\gt(z)}$ of $A$ is maximal dissipative in $\sH$. 
%This family is called the {\it \v{S}traus family of $A$ associated with $\tau(\cdot)$}.

Under additional assumptions ${\gt(\cdot)}$ 
can be realized as the Weyl function corresponding to a densely defined closed simple symmetric
operator $T$ with finite deficiency indices in some Hilbert space $\gotG$ and a boundary triplet 
$\gP_T = \{\kH,\gY_0,\gY_1\}$ for $T^*$. Then the selfadjoint (exit space) extension $\widetilde L$ of $A$ 
can be recovered as a coupling of the operators $A$ and $T$ corresponding to a coupling of the boundary 
triplets $\Pi_A$ and $\Pi_T$ (see \cite{DHMS00} and formula
\eqref{athe} below). We prove in Theorem~\ref{t.IV.2} that for such systems there exists a
spectral shift function $\wt \xi(\cdot)$ given by
\bed
\widetilde{\xi}(\gl) =
\frac{1}{\pi}\imag\bigl(\tr\bigl(\log(M(\gl+i0) + \gt(\gl+i0))\bigr)\bigr)
\eed
and that the modified trace formula
\bed
\begin{split}
&\tr\left((A_{-\gt(z)} - z)^{-1} - (A_0 - z)^{-1}\right) +\\
&\qquad
\tr\left((T_{-M(z)} - z)^{-1} - (T_0 - z)^{-1}\right) =
-\int_\bR \frac{1}{(\gl - z)^{2}}\,\widetilde{\xi}(\gl)\, d\gl
\end{split}
\eed
holds for all $z \in \bC \setminus \bR$. 
%Setting $L_0:= A_0\oplus T_0$, where $T_0:=
%T^*\upharpoonright\ker(\gY_0)$ and assuming that $T$ has finite
%deficiency indices formula \eqref{0.7}  applies to  the scattering system $\{\wt L, L_0\}$.
%This leads to the same formula for the scattering matrix $\wt S(\cdot) := \wt S(\wt L,L_0;\cdot)$
%with $\wt M(\cdot) := \diag(M(\cdot),\gt(\cdot) )$
%instead of $M(\cdot)$ and $\Theta$ given by \eqref{wttheta}.

Let $T_0=T^*\upharpoonright\ker(\gY_0)$ be the selfadjoint extension of $T$ in $\gotG$ corresponding to the
boundary mapping $\gY_0$. With the help of the channel wave operators
\begin{equation*}
\begin{split}
W_\pm(\wt L,A_0) &=
s-\lim_{t\to\pm\infty}e^{it\wt L}e^{-itA_0}P^{ac}(A_0)\\
W_\pm(\wt L,T_0) &= s-\lim_{t\to\pm\infty}e^{it\wt L}e^{-itT_0}P^{ac}(T_0)
\end{split}
\end{equation*}
one then defines the channel scattering operators 
\begin{equation*}
S_\gotH := W_+(\widetilde{L},A_0)^*W_-(\widetilde{L},A_0)
\quad \mbox{and} \quad
S_\gotG :=  W_+(\widetilde{L},T_0)^*W_-(\widetilde{L},T_0).
\end{equation*}
The corresponding channel scattering matrices $S_\gotH(\gl)$ and
$S_\gotG(\gl)$ are studied in Section~\ref{4.3sec}. Here we express these scattering matrices 
in terms of the functions $M$ and $\tau$ in the spectral representations 
$L^2(\dR,d\lambda,\cH_{M(\lambda)})$ and $L^2(\dR,d\lambda,\cH_{\tau(\lambda)})$ of $A_0^{ac}$ and $T_0^{ac}$, respectively,
and finally, with the help of these
representations the modified Birman-Krein formula
\begin{equation*}
\det(S_\gotH(\gl)) = \overline{\det(S_\gotG(\gl))}
\exp\bigl(-2\pi i\widetilde{\xi}(\gl)\bigr)
\end{equation*}
is proved in Theorem~\ref{t.IV.5}.

% We note the interesting relation
%$\wt\xi(\gl) = \eta_{\gt(\gl)}(\gl)$ for a.e. $\gl \in \bR$.

\section{Self-adjoint extensions and scattering}\label{two}

In this section we consider scattering systems consisting of 
two selfadjoint extensions of a densely defined
symmetric operator with equal finite deficiency indices in a separable
Hilbert space. We generalize a result on the representation
of the spectral shift function of such a scattering system from
\cite{BeMN1} and we give a short proof of the Birman-Krein formula in this setting.

\subsection{Boundary triplets and closed extensions}\label{btrips}

Let $A$ be a densely defined closed symmetric operator with equal
deficiency indices $n_\pm(A)=\dim\ker(A^*\mp i)\leq\infty$ in the
separable Hilbert space $\gotH$. We use a  concept of a boundary
triplet for $A^*$ in order to describe of the closed extensions
$A_\gT\subset A^*$ of $A$ in $\gotH$, see \cite{GG} and also \cite{DM91,DM95}.

\begin{defn}
A triplet $\Pi=\{\kH,\gG_0,\gG_1\}$ is called a {\rm boundary triplet} for the adjoint
operator $A^*$ if $\kH$ is a Hilbert space and
$\Gamma_0,\Gamma_1:\  \dom(A^*)\rightarrow\kH$ are linear mappings such that

\begin{enumerate}

\item [{\rm (i)}] the abstract second  Green's  identity,
\begin{equation*}
(A^*f,g) - (f,A^*g) = (\gG_1f,\gG_0g) - (\gG_0f,\gG_1g),
\end{equation*}
holds for all $f,g\in\dom(A^*)$ and

\item [{\rm (ii)}] the mapping
$\gG:=(\Gamma_0,\Gamma_1)^\top:  \dom(A^*) \longrightarrow \kH
\times \kH$ is surjective.
\end{enumerate}
\end{defn}

We refer to \cite{DM91} and \cite{DM95} for a detailed study of
boundary triplets and recall only some important facts. First of all
a boundary triplet $\Pi=\{\kH,\gG_0,\gG_1\}$ for $A^*$ exists since
the deficiency indices $n_\pm(A)$ of $A$ are assumed to be equal.
Then $n_\pm(A) = \dim\kH$ holds. We note that a boundary triplet for
$A^*$ is not unique. Namely, if
$\Pi^\prime=\{\cG^\prime,\Gamma_0^\prime,\Gamma_1^\prime\}$ is another boundary triplet for
$A^*$, then there exists a boundedly invertible operator $W=(W_{ij})_{i,j=1}^2\in[\cG\oplus\cG,\cG^\prime\oplus\cG^\prime]$
with the property
\begin{equation*}
W^* \begin{pmatrix}0 & -iI_{\cG^\prime}\\ iI_{\cG^\prime} & 0\end{pmatrix}W
= \begin{pmatrix}0 & -iI_{\cG}\\ iI_{\cG} & 0\end{pmatrix},
\end{equation*}
such that
\begin{equation*}
\begin{pmatrix}\Gamma_0^\prime\\\Gamma_1^\prime\end{pmatrix}
=\begin{pmatrix}W_{11} & W_{12}\\ W_{21} & W_{22}\end{pmatrix}
\begin{pmatrix}\Gamma_0\\\Gamma_1\end{pmatrix}
\end{equation*}
holds.
%%, so that formally $W=\Gamma^\prime\circ\Gamma^{-1}$. 
Here and in the following we write $[\sK,\cK]$ for the set of bounded everywhere defined linear operators
acting from a Hilbert space $\sK$ into a Hilbert space $\cK$. For brevity we write $[\cK]$ if $\cK=\sK$.

An operator $A^\prime$ is called a {\it proper extension} of $A$
if $A^\prime$ is closed and satisfies $A\subseteq A^\prime \subseteq A^*$. 
In order to describe the set of proper extensions
of $A$ with the help of a boundary triplet
$\Pi=\{\kH,\Gamma_0,\Gamma_1\}$ for $A^*$ we have to consider the
set $\widetilde\kC(\kH)$ of closed linear relations in $\kH$, that
is, the set of closed linear subspaces of $\kH\times\kH$. Linear operators in $\kH$ are identified with 
their graphs, so that the
set $\kC(\kH)$ of closed linear operators 
is viewed as a subset of $\widetilde\kC(\kH)$. For the usual definitions of the
linear operations with linear relations, the inverse, the resolvent
set and the spectrum we refer to \cite{DS87}. Recall that the
adjoint relation $\gT^*\in\widetilde\kC(\kH)$ of a linear
relation $\gT$ in $\kH$ is defined as
\begin{equation}\label{thetastar}
\gT^*:= \left\{ \begin{pmatrix} k\\ k^\prime\end{pmatrix}:
(h^\prime,k)=(h,k^\prime)\,\,\text{for all}\,\,
\begin{pmatrix} h\\ h^\prime\end{pmatrix} \in\gT\right\}
\end{equation}
and $\gT$ is said to be {\it symmetric} ({\it selfadjoint}) if
$\gT\subseteq\gT^*$ (resp. $\gT=\gT^*$). Note that
definition \eqref{thetastar} extends the definition of the adjoint operator.
A linear relation $\Theta$ is called {\it dissipative} 
if $\imag(g^\prime,g)\leq 0$ holds for all $\bigl(\begin{smallmatrix} g\\ g^\prime\end{smallmatrix}\bigr)\in\Theta$ and $\Theta$
is said to be {\it maximal dissipative} if $\Theta$ is dissipative and each dissipative extension of $\Theta$ coincides with $\Theta$ itself.
In this case the upper half plane $\dC_+=\{\lambda\in\dC:\imag\lambda>0\}$ belongs to the resolvent set $\rho(\Theta)$.
Furthermore, a linear relation $\Theta$ is called {\it accumulative} ({\it maximal accumulative}) if $-\Theta$ is dissipative (resp. maximal dissipative).
For a maximal accumulative relation $\Theta$ we have $\dC_-=\{\lambda\in\dC:\imag\lambda<0\}\subset\rho(\Theta)$.

With a  boundary triplet  $\Pi=\{\kH,\gG_0,\gG_1\}$ for $A^*$ one associates
two selfadjoint extensions of $A$ defined by
\begin{equation*}
A_0:=A^*\!\upharpoonright\ker(\gG_0)
\quad \text{and}\quad
A_1:=A^*\!\upharpoonright\ker(\gG_1).
\end{equation*}
A description of all proper extensions
of $A$ is given in the next proposition.
Note also that the selfadjointness of $A_0$ and $A_1$ is a
consequence of Proposition \ref{propo} (ii).

\begin{prop}\label{propo}
Let $A$ be a densely defined closed symmetric operator in $\sH$ and let
$\Pi=\{\kH,\gG_0,\gG_1\}$ be a  boundary triplet for $A^*$.  Then the mapping
\begin{equation}\label{bij}
\gT\mapsto A_\gT:= \Gamma^{-1}\gT=\bigl\{f\in\dom(A^*): \
\Gamma f=(\Gamma_0 f,\Gamma_1 f)^\top \in \gT\bigr\}
\end{equation}
establishes  a bijective correspondence between the set
$\widetilde\kC(\kH)$ and the set of proper extensions of $A$.
Moreover, for $\gT\in\widetilde\kC(\kH)$ the following assertions
hold.

\begin{enumerate}

\item [{\rm (i)}] $(A_\gT)^*=  A_{\gT^*}$.

\item [{\rm (ii)}] $A_\gT$ is  symmetric (selfadjoint) if and only if $\gT$ is
symmetric (resp. selfadjoint).

\item [{\rm (iii)}] $A_\gT$ is  dissipative (maximal dissipative) if and only if $\gT$ is
dissipative (resp. maximal dissipative).

\item [{\rm (iv)}] $A_\gT$ is accumulative (maximal accumulative) if and only if $\gT$ is
accumulative (resp. maximal accumulative).

\item [{\rm (v)}]
$A_\gT$ is disjoint with $A_0,$ that is $\dom(A_\gT)\cap \dom(A_0) =\dom(A),$ if
and only if $\gT\in \kC(\kH)$. In this case
the extension $A_\gT$ in \eqref{bij} is given by
\begin{equation}\label{0}
A_\gT=A^*\!\upharpoonright
\ker\bigl(\Gamma_1-\gT\Gamma_0\bigr).
\end{equation}
\end{enumerate}
\end{prop}

We note that \eqref{0} holds also for linear relations $\Theta$ if the expression 
$\Gamma_1-\Theta\Gamma_0$ is interpreted in the sense of linear relations.

In the following we shall often be concerned with simple symmetric
operators. Recall that a symmetric operator is said to be {\it
simple} if there is no nontrivial subspace which reduces it to a
selfadjoint operator. By \cite{K49} each symmetric operator $A$ in
$\gotH$ can be written as the direct orthogonal sum $\widehat A\oplus
A_s$ of a simple symmetric operator $\widehat A$ in the Hilbert
space
\begin{equation*}
\widehat\gotH=\clo\spa\bigl\{\ker(A^*-\gl):
\gl\in\bC\backslash\bR\bigr\}
\end{equation*}
and a selfadjoint operator $A_s$ in $\gotH\ominus\widehat\gotH$. Here
$\clospa \{\cdot\}$ denotes the closed linear span of a set.
Obviously $A$ is simple if and only if $\widehat\gotH$ coincides
with $\gotH$.

\subsection{Weyl functions and resolvents of extensions}\label{weylreso}

Let, as in Section \ref{btrips}, $A$ be a densely defined closed
symmetric operator in $\gotH$ with equal deficiency indices. If
$\lambda\in\dC$ is a point of regular type of $A$, i.e.
$(A-\lambda)^{-1}$ is bounded, we denote the {\it defect subspace}
of $A$ by $\kN_\gl=\ker(A^*-\gl)$. The following definition can 
be found in \cite{DM87,DM91,DM95}.

\begin{defn}\label{Weylfunc}
Let $\Pi=\{\kH,\gG_0,\gG_1\}$ be a boundary triplet for the operator $A^*$ and
let $A_0=A^*\!\upharpoonright\ker(\gG_0)$. The operator-valued
functions
$\gamma(\cdot) :\ \rho(A_0)\rightarrow  [\kH,\gotH]$ and  $M(\cdot) :\
\rho(A_0)\rightarrow  [\kH]$ defined by
\begin{equation}\label{2.3A}
\gamma(\gl):=\bigl(\Gamma_0\!\upharpoonright\kN_\gl\bigr)^{-1} \qquad\text{and}\qquad
M(\gl):=\Gamma_1\gamma(\gl), \quad \gl\in\rho(A_0),
\end{equation}
are called the {\rm $\gamma$-field} and the {\rm Weyl function}, respectively,
corresponding to the boundary triplet $\Pi=\{\kH,\gG_0,\gG_1\}$.
\end{defn}

It follows from the identity  $\dom(A^*)=\ker(\Gamma_0)\,\dot
+\,\kN_\gl$, $\lambda\in\rho(A_0)$, where as above
$A_0=A^*\!\upharpoonright\ker(\gG_0)$, that the $\gamma$-field
$\gamma(\cdot)$ in \eqref{2.3A} is well defined. It is easily seen
that both  $\gamma(\cdot)$ and $M(\cdot)$ are holomorphic on
$\rho(A_0)$ and the relations
\begin{equation}\label{gammamu}
\gamma(\mu)=\bigl(I+(\mu-\gl)(A_0-\mu)^{-1}\bigr)\gamma(\gl),
\qquad \gl,\mu\in\rho(A_0),
\end{equation}
and
\begin{equation}\label{mlambda}
M(\gl)-M(\mu)^*=(\gl-\bar\mu)\gamma(\mu)^*\gamma(\gl),
\qquad \gl,\mu\in\rho(A_0),
\end{equation}
are valid (see \cite{DM91}).
The identity \eqref{mlambda} yields that $M(\cdot)$ is a  {\it
Nevanlinna function}, that is, $M(\cdot)$ is holomorphic on $\bC\backslash\bR$ and
takes values in $[\kH]$, $M(\gl)=M(\overline\gl)^*$ for
all $\gl\in\bC\backslash\bR$ and $\imag(M(\gl))$ is a nonnegative operator
for all $\gl$ in the upper half plane $\bC_+$.
Moreover, it follows 
that $0\in \rho(\imag(M(\gl)))$ holds.
It is important to note that if the operator $A$ is simple, then the
Weyl function $M(\cdot)$ determines the pair $\{A,A_0\}$ uniquely up
to unitary equivalence, cf. \cite{DM87,DM91}.

In the case that the deficiency indices $n_+(A)=n_-(A)$ are finite
the Weyl function $M$ corresponding to the boundary triplet $\Pi=\{\cH,\Gamma_0,\Gamma_1\}$
is a matrix-valued Nevanlinna function in the finite dimensional space
$\cH$. From \cite{Don1,Gar1} one gets  the existence of the (strong) limit
\begin{equation*}
M(\lambda+i0)=\lim_{\epsilon\rightarrow +0} M(\lambda+i\epsilon)
\end{equation*}
from the upper half plane for a.e. $\lambda\in\dR$.

Let now $\Pi=\{\kH,\Gamma_0,\Gamma_1\}$ be a boundary triplet for
$A^*$ with $\gamma$-field $\gamma(\cdot)$ and Weyl function
$M(\cdot)$. The spectrum and the resolvent set of a proper (not
necessarily selfadjoint) extension of $A$ can be described with the
help of the Weyl function. If $A_\gT\subseteq A^*$ is the
extension corresponding to $\gT\in\widetilde\kC(\kH)$ via
\eqref{bij}, then a point $\gl\in\rho(A_0)$
belongs to $\rho(A_\gT)$ ($\sigma_i(A_0)$, $i=p,c,r$)
if and only if $0\in\rho(\gT-M(\gl))$ (resp.
$0\in\sigma_i(\gT-M(\gl))$, $i=p,c,r$). Moreover, for
$\gl\in\rho(A_0)\cap\rho(A_\gT)$ the well-known resolvent formula
\begin{equation}\label{resoll}
(A_\gT - \gl)^{-1} = (A_0 - \gl)^{-1} + \gga(\gl)\bigl(\gT -
M(\gl)\bigr)^{-1}\gga(\bar{\gl})^*
\end{equation}
holds. Formula \eqref{resoll} is a generalization of the known Krein
formula for canonical resolvents. We emphasize that it is valid for
any proper extension of $A$ with a non-empty resolvent set. 
It is worth to note that the Weyl function can also be used to investigate the
absolutely continuous and singular continuous spectrum of proper extensions
of $A$, cf. \cite{BMN1}.

\subsection{Spectral shift function and trace formula}

M.G.~Krein's spectral shift function introduced in \cite{K62} is an important tool in
the spectral and perturbation theory of selfadjoint operators, in particular 
scattering theory. A detailed
review on the spectral shift function can be
found in, e.g. \cite{BY92a,BY92b}. Furthermore we mention \cite{GMN1,GM1,GM2}
as some recent papers on the spectral shift function and its various applications. 

Recall that for any pair of selfadjoint operators $H_1,H_0$ in
a separable Hilbert space $\gotH$ such that the resolvents differ by
a trace class operator,
\begin{equation}\label{trace1}
(H_1-\gl)^{-1}-(H_0-\gl)^{-1}\in\gotS_1(\gotH),
\end{equation}
for some (and hence for all) $\lambda\in\rho(H_1)\cap\rho(H_0)$,
there exists a real valued function $\xi(\cdot)\in L^1_{loc}(\bR)$
satisfying the conditions
\begin{equation}\label{shift1}
\tr\left((H_1-\gl)^{-1}-(H_0-\gl)^{-1}\right)= -\int_{\bR}
\frac{1}{(t-\gl)^{2}}\,\xi(t)\,dt,
\end{equation} 
$\gl\in\rho(H_1)\cap\rho(H_0)$, and
\begin{equation}\label{shift2}
\int_{\bR} \frac{1}{1+t^2}\,\xi(t)\, dt <\infty,
\end{equation}
cf. \cite{BY92a,BY92b,K62}.
Such a function $\xi$ is called a {\it spectral shift
function} of the pair $\{H_1,H_0\}$. We emphasize that $\xi$
is not unique, since simultaneously with $\xi$ a function
$\xi+c$, $c\in\bR$, also satisfies both conditions
\eqref{shift1} and \eqref{shift2}. Note that the converse also
holds, namely, any two spectral shift functions for a pair of selfadjoint
operators $\{H_1,H_0\}$ satisfying \eqref{trace1} 
differ by a real constant. We remark that \eqref{shift1} is a
special case of the general formula
\begin{equation}\label{shift1A}
\tr\left(\phi(H_1) - \phi(H_0)\right)= \int_{\bR}
\phi'(t)\,\xi(t)\,dt,
\end{equation}
which is valid for a wide class of smooth functions, cf. \cite{Pe1} for a large
class of such functions $\phi(\cdot)$.

In Theorem~\ref{V.1} below we find a representation for the spectral shift
function $\xi_\gT$
of a pair of selfadjoint operators $A_\gT$ and $A_0$ which are both assumed to be
extensions of a densely defined
closed simple symmetric operator $A$ with equal finite deficiency
indices. For that purpose we use the definition
\begin{equation}\label{log}
\log (T):=-i\int_0^\infty \bigl((T+it)^{-1}-(1+it)^{-1}I_\cH\bigr)\,dt
\end{equation}
for an operator $T$ in a finite dimensional Hilbert space $\cH$ satisfying
$\imag(T)\geq 0$ and $0\not\in\sigma(T)$, see, e.g. \cite{GMN1,Pot}. 
A straightforward calculation shows that
the relation 
\begin{equation}\label{det}
\det(T)=\exp\bigl(\tr\bigl(\log(T)\bigr)\bigr)
\end{equation}
holds. Observe that
\begin{equation}\label{detlog}
\tr\bigl(\log(T)\bigr)=\log\bigl(\det (T)\bigr)+2k\pi i
\end{equation}
holds for some $k\in\dZ$.
In \cite[Theorem 4.1]{BeMN1} it was shown that if $\Pi=\{\cH,\Gamma_0,\Gamma_1\}$ is a boundary triplet
for $A^*$ with $A_0=A^*\upharpoonright\ker(\Gamma_0)$ and $A_\Theta=A^*\upharpoonright \ker(\Gamma_1-\Theta\Gamma_0)$
is a selfadjoint extension of $A$ which corresponds to a selfadjoint matrix $\Theta$ in $\cH$, then the limit
$\lim_{\epsilon\to+0}\log\left(M(\gl + i\epsilon) - \gT\right)$ exist for a.e. $\gl
\in \bR$ and 
\begin{equation}\la{t.2.13}
\xi_\gT(\gl) := \frac{1}{\pi}\imag\bigl(\tr\bigl(\log(M(\gl + i0) -\gT)\bigr)\bigr)
\end{equation}
defines a spectral shift function for the pair $\{A_\gT,A_0\}$. We emphasize that $\Theta$ was assumed to be a
matrix in \cite{BeMN1}, so that $\xi_\gT$ in \eqref{t.2.13} is a spectral shift function only for special pairs $\{A_\gT,A_0\}$.
Theorem~\ref{V.1} below extends the result from \cite{BeMN1}  to the case of a selfadjoint relation $\gT$ and hence to arbitrary pairs of
selfadjoint extensions $\{A_\Theta,A_0\}$ of $A$.

To this end we first recall that any selfadjoint relation $\gT$ in $\cH$
can be written in the form 
\begin{equation}\la{t.2.14}
\gT=\gT_{\rm op}\oplus\gT_{\infty}
\end{equation}
with respect to the decomposition $\cH=\cH_\op\oplus\cH_\infty$, where $\gT_{\rm op}$ is
a selfadjoint operator in $\cH_\op:=\overline{\dom\gT}$ and 
$\gT_{\infty}$ is a pure relation in $\cH_\infty:=(\dom\Theta)^\perp$, that is,
\begin{equation}\la{t.2.15}
\gT_\infty =\left\{
\begin{pmatrix}
0\\
h^\prime
\end{pmatrix}: h^\prime\in\cH_\infty\right\}.
\end{equation}
Since in the following considerations the space $\cH$ is finite dimensional we have $\cH_\op=\dom\Theta=\dom\Theta_\op$ and $\Theta_\op$ is a selfadjoint matrix.
If $M(\cdot)$ is the Weyl function corresponding to a boundary triplet $\Pi=\{\cH,\Gamma_0,\Gamma_1\}$, then 
\begin{equation}\label{mop}
M_{\rm op}(\gl)  := P_{\rm op}M(\gl)\iota_\op,
\end{equation}
is a $[\cH_\op]$-valued Nevanlinna function. Here $P_{\rm op}$ is the orthogonal projection from  $\cH$ onto $\cH_\op$ 
and $\iota_\op$ denotes  the canonical embedding of $\cH_\op$ in $\cH$.
One verifies that
\begin{equation}\la{t.2.5}
\bigl(\gT - M(\gl)\bigr)^{-1} = \iota_\op\bigl(\gT_{\rm op} - M_{\rm op}(\gl)\bigr)^{-1}P_{\rm op}
\end{equation}
holds for all $\gl \in \bC_+$. The following result generalizes \cite[Theorem 4.1]{BeMN1}, see also \cite{LSY} for a special case.
\begin{thm}\label{V.1}
Let $A$ be a densely defined closed simple symmetric operator in the separable Hilbert space
$\gotH$ with equal finite deficiency indices, let $\Pi=
\{\kH,\Gamma_0,\Gamma_1\}$ be a boundary triplet for $A^*$ and let
$M(\cdot)$ be the corresponding Weyl function. Furthermore, let
$A_0=A^*\!\upharpoonright\ker(\Gamma_0)$  and let
$A_\gT=A^*\upharpoonright\Gamma^{-1}\gT$, $\Theta\in\widetilde\cC(\cH)$,
be a selfadjoint extension of $A$ in $\sH$. Then 
the limit 
\begin{equation*}
\lim_{\epsilon\rightarrow +0}\log\bigl(M_{\rm op}(\lambda+i\epsilon)-\gT_{\rm op}\bigr)
\end{equation*}
exists for a.e. $\lambda\in\dR$ and the function
\begin{equation}\label{xitheta}
\xi_\gT(\lambda):=
\frac{1}{\pi}\imag\bigl(\tr\bigl(\log(M_{\rm op}(\gl + i0) - \gT_{\rm op})\bigr)\bigr) 
\end{equation}
is a spectral shift function for the pair $\{A_\gT,A_0\}$ with
$0\leq\xi_\gT(\gl) \le \dim\cH_\op$.
\end{thm}
\begin{proof}
Since $\lambda\mapsto M_{\rm op}(\lambda)-\gT_{\rm op}$ 
is a Nevanlinna function with values in $[\cH_\op]$ 
and $0\in\rho(\imag(M_{\rm op}(\lambda)))$ for all $\lambda\in\dC_+$, it
follows that $\log(M_{\rm op}(\lambda)-\gT_{\rm op})$ is
well-defined for all  $\lambda\in\dC_+$ by \eqref{log}.
According to \cite[Lemma 2.8]{GMN1} the function $\lambda\mapsto
\log(M_{\rm op}(\lambda)-\gT_{\rm op})$, 
$\lambda\in\dC_+$, is a $[\cH_\op]$-valued Nevanlinna function such that
\begin{equation*}
0\leq\imag\bigl(\log(M_{\rm op}(\lambda)-\gT_{\rm op})\bigr)\leq \pi I_{\cH_\op}
\end{equation*}
holds for all $\lambda\in\dC_+$. Hence the limit $\lim_{\epsilon\rightarrow +0}
\log(M_{\rm op}(\lambda+i\epsilon)-\gT_{\rm op})$ 
exists for a.e. $\lambda\in\dR$ (see \cite{Don1,Gar1} and
Section~\ref{weylreso}) and 
$\lambda\mapsto\tr(\log(M_{\rm op}(\lambda)-\gT_{\rm op}))$,
$\lambda\in\dC_+$, is a scalar Nevanlinna
function with the property 
\begin{equation*}
0\leq\imag\bigl(\tr(\log(M_{\rm op}(\lambda)-\gT_{\rm op}))\bigr)\leq \pi\dim\cH_\op,\quad\lambda\in\dC_+,
\end{equation*}
that is, the function $\xi_\gT$ in \eqref{xitheta} satisfies $0\leq\xi_\gT(\gl) \le \dim\cH_\op$ for 
a.e. $\lambda\in\dR$.

In order to show that \eqref{shift1} holds with $H_1$, $H_0$ and $\xi$ replaced by
$A_\gT$, $A_0$ and $\xi_\gT$, respectively, we note  
that the relation
\begin{equation}\label{trlog}
\frac{d}{d\lambda}\tr\bigl(\log(M_{\rm op}(\lambda)-\gT_{\rm op})\bigr)=
\tr\left((M_\op(\lambda)-\gT_\op)^{-1}
\frac{d}{d\lambda}M_\op(\lambda)\right) 
\end{equation}
is true for all $\lambda\in\dC_+$. This can be shown in the same way as in the proof of \cite[Theorem 4.1]{BeMN1}. 
From \eqref{mlambda} we find
\begin{equation}\label{4.255}
\gamma(\bar\mu)^*\gamma(\gl)=\frac{M(\gl)-M(\bar\mu)^*} {\gl-\mu}, \qquad \gl,
\mu\in\dC\backslash\dR,\,\lambda\not=\mu,
\end{equation}
and passing in \eqref{4.255} to the limit $\mu\to\gl$ one gets
\begin{equation}\la{5.9}
\gamma(\bar\gl)^*\gamma(\gl)=\frac{d}{d\gl}M(\gl).
\end{equation}
Making use of formula \eqref{resoll} for the canonical resolvents this implies
\begin{equation}\label{5.10}
\begin{split}
\tr\left((A_\gT-\gl)^{-1}-(A_0-\gl)^{-1}\right)&=
-\tr\left((M(\gl) - \gT)^{-1}\gamma(\bar\gl)^*\gamma(\gl)\right)\\
&=-\tr\left( (M(\gl) - \gT)^{-1}   \frac{d}{d\gl}M(\gl)\right)
\end{split}
\end{equation}
for all $\gl \in \bC_+$. With respect to the decomposition $\cH=\cH_\op\oplus\cH_\infty$ the operator
\begin{equation*}
\bigl(M(\gl) - \gT\bigr)^{-1}   \frac{d}{d\gl}M(\gl)=\iota_{\cH_\op}\bigl(M_\op(\lambda)-\Theta_\op\bigr)^{-1}P_\op\frac{d}{d\lambda} M(\lambda)
\end{equation*}
is a $2\times 2$ block matrix where the entries in the lower row are zero matrices and the upper left corner
is given by
\begin{equation*}
(M_\op(\lambda)-\Theta_\op)^{-1}\frac{d}{d\lambda} M_\op(\lambda).
\end{equation*}
Therefore \eqref{5.10} becomes
\begin{equation}\label{5.111}
\begin{split}
\tr\left((A_\gT-\gl)^{-1}-(A_0-\gl)^{-1}\right)
&=-\tr\left( (M_\op(\gl) - \gT_\op)^{-1}   \frac{d}{d\gl}M_\op(\gl)\right),\\
&=-\frac{d}{d\lambda}\tr\bigl(\log(M_{\rm op}(\lambda)-\gT_{\rm op})\bigr),
\end{split}
\end{equation}
where we have use \eqref{trlog}.

Further, by \cite[Theorem 2.10]{GMN1} there exists a $[\cH_\op]$-valued
measurable function $t\mapsto \Xi_{\gT_{\rm op}}(t)$, $t\in\dR$, such that
$\Xi_{\gT_{\rm op}}(t)=\Xi_{\gT_\op}(t)^*$ and
$0 \le \Xi_{\gT_\op}(t) \le I_{\kH_\op}$
for a.e. $\gl \in \bR$ and the representation
\bed
\log(M_{\rm op}(\gl) - \gT_{\rm op}) = 
C + \int_\bR \Xi_{\gT_{\rm op}}(t)\left((t - \gl)^{-1} - t(1 + t^2)^{-1}\right)\,dt,
\quad \gl \in \bC_+,
\eed
holds with some bounded selfadjoint
operator $C$. Hence
\bed
\tr\bigl(\log(M_{\rm op}(\gl) - \gT_{\rm op})\bigr) =
\tr(C) + \int_\bR  \tr\left(\Xi_{\gT_{\rm op}}(t)\right)
\left((t - \gl)^{-1} - t(1 + t^2)^{-1}\right)\,dt
\eed
for $\gl \in \bC_+$ and we conclude from
\begin{equation*}
\begin{split}
\xi_\gT(\lambda)&=\lim_{\epsilon\rightarrow +0}
\frac{1}{\pi}\imag\bigl(\tr(\log(M_{\rm op}(\gl + i\epsilon) -
\gT_{\rm op}))\bigr)\\
&=\lim_{\epsilon\rightarrow
  +0}\frac{1}{\pi}\int_\dR\tr\left(\Xi_{\gT_\op}(t)\right)
\epsilon\bigl((t-\lambda)^2+\epsilon^2\bigr)^{-1}dt 
\end{split}
\end{equation*}
that $\xi_\gT(\gl) = \tr(\Xi_{\gT_{\rm op}}(\gl))$
is true for a.e. $\gl \in \bR$. Therefore we have 
\begin{equation*}
\frac{d}{d\gl}\tr\bigl(\log(M_{\rm op}(\gl) - \gT_{\rm op})\bigr) =
\int_\bR  (t - \gl)^{-2}\xi_\gT(t)\,dt
\end{equation*}
and together with \eqref{5.111} we
immediately get the trace formula 
\begin{equation*}
\tr\left((A_\gT-\gl)^{-1}-(A_0-\gl)^{-1}\right)= -\int_{\bR}
\frac{1}{(t-\gl)^{2}}\,\xi_\gT(t)\,dt.
\end{equation*}
The integrability condition \eqref{shift2} holds because of \cite[Theorem~2.10]{GMN1}.
This completes the proof of Theorem~\ref{V.1}.
\end{proof}

\subsection{A representation of the scattering matrix}\label{scat}

Let again $A$ be a densely defined closed simple symmetric operator in the separable Hilbert space
$\gotH$ with equal finite deficiency indices and let  $\Pi=\{\kH,\gG_0,\gG_1\}$
be a boundary triplet for $A^*$ with $A_0=A^*\upharpoonright\ker(\Gamma_0)$.
Let $\Theta$ be a selfadjoint relation in $\kH$ 
and let $A_\Theta=A^*\upharpoonright\Gamma^{-1}\Theta$ be the corresponding selfadjoint extension of $A$ in $\gotH$.
Since $\dim\cH$ is finite by \eqref{resoll}
\begin{equation*}
\dim\Bigl(\ran\bigl((A_\Theta-\lambda)^{-1}-(A_0-\lambda)^{-1}\bigr)\Bigr)
<\infty,\quad\lambda\in\rho(A_\Theta)\cap\rho(A_0),
\end{equation*}
and therefore the pair $\{A_\gT,A_0\}$ forms a so-called
{\it complete scattering system}, that is, the {\it wave operators}
\begin{equation*}
W_\pm(A_\Theta,A_0) := \slim_{t\to\pm\infty}e^{itA_\Theta}e^{-itA_0}P^{ac}(A_0),
\end{equation*}
exist and their ranges coincide with the absolutely continuous
subspace $\gotH^{ac}(A_\Theta)$ of $A_\Theta$, cf. \cite{BW,Ka1,Wei1,Y}.
 $P^{ac}(A_0)$ denotes the orthogonal
projection onto the absolutely continuous subspace $\gotH^{ac}(A_0)$
of $A_0$.
The {\it scattering operator} $S_\Theta$ of the {\it scattering system}
$\{A_\Theta,A_0\}$ is then defined by
\begin{equation*}
S_\Theta:= W_+(A_\Theta,A_0)^*W_-(A_\Theta,A_0).
\end{equation*}
If we regard the scattering operator as an operator in $\gotH^{ac}(A_0)$,
then $S_\Theta$ is unitary, commutes with the absolutely continuous part
\begin{equation*}
A^{ac}_0:=A_0\upharpoonright \dom(A_0)\cap\gotH^{ac}(A_0)
\end{equation*}
of $A_0$ and it follows
that $S_\Theta$ is unitarily equivalent to a multiplication operator
induced by a family $\{S_\Theta(\lambda)\}_{\lambda\in\dR}$ of unitary operators in
a spectral representation of $A_0^{ac}$, see e.g. \cite[Proposition 9.57]{BW}.
This family is called
the {\it scattering matrix} of the scattering system $\{A_\Theta,A_0\}$. 

We recall a representation theorem for  the scattering matrix 
$\{S_\Theta(\lambda)\}_{\lambda\in\dR}$ in terms of the Weyl
function $M(\cdot)$ of the boundary triplet $\Pi=\{\kH,\Gamma_0,\Gamma_1\}$ from \cite{BeMN1}.
For this we consider the Hilbert space $L^2(\bR,d\lambda,\kH)$, where $d\gl$
is the Lebesgue measure on $\bR$. Further, we set
\begin{equation}\la{2.7}
\kH_{M(\gl)} := \ran\bigl(\imag(M(\gl))\bigr), \quad M(\gl) := M(\gl + i0),
\end{equation}
which defines subspaces of $\kH$ for a.e. $\gl \in \bR$. By $P_{M(\gl)}$ we denote
the orthogonal projection from $\kH$ onto $\kH_{M(\gl)}$. The family
$\{P_{M(\gl)}\}_{\gl \in \bR}$ is measurable. Hence $\{P_{M(\gl)}\}_{\gl \in
  \bR}$ induces a multiplication operator $P_M$ on $L^2(\bR,d\gl,\kH)$
defined by
\begin{equation}
(P_Mf)(\gl ) =  P_{M(\gl)}f(\gl), \qquad f \in L^2(\bR,d\lambda,\kH), 
\end{equation}
which is an orthogonal projection. The subspace $\ran(P_M)$ is
denoted by $L^2(\bR,d\lambda,\kH_{M(\gl)})$ in the following. We remark that $L^2(\bR,d\lambda,\kH_{M(\gl)})$ can be
regarded as the direct integral of the Hilbert spaces $\kH_{M(\gl)}$, that
is, 
\begin{equation*}
L^2(\bR,d\gl,\kH_{M(\gl)}) = \int^\oplus \kH_{M(\gl)}\,d\lambda.
\end{equation*} 
The following theorem was proved in  \cite{BeMN1}.
\begin{thm}\label{scattering}
Let $A$ be a densely defined closed simple symmetric operator with
equal finite deficiency indices in the separable Hilbert space $\gotH$ and
let $\Pi= \{\kH,\Gamma_0,\Gamma_1\}$ be a boundary triplet for $A^*$
with corresponding Weyl function $M(\cdot)$. Furthermore, let
$A_0=A^*\!\upharpoonright\ker(\Gamma_0)$  and
let $A_\gT=A^*\upharpoonright\Gamma^{-1}\gT$,
$\gT\in\widetilde\kC(\kH)$, be a selfadjoint extension of $A$ in $\sH$.
Then the following holds:
\begin{itemize}
 \item [{\rm (i)}] $A_0^{ac}$ is unitarily equivalent to the multiplication operator with the free variable in the Hilbert space $L^2(\dR,d\lambda,\kH_{M(\gl)})$.
 \item [{\rm (ii)}] In the spectral representation $L^2(\bR,d\gl,\kH_{M(\gl)})$ of
$A^{ac}_0$ the scattering matrix $\{S_\gT(\gl)\}_{\gl
\in \bR}$ of the scattering system $\{A_\gT,A_0\}$ admits the
representation
\begin{equation}\label{scatformula}
S_\gT(\gl) = I_{\kH_{M(\gl)}} +
2i\sqrt{\imag(M(\gl))}\bigl(\gT-M(\gl)\bigr)^{-1} \sqrt{\imag(M(\gl))}
\end{equation}
for a.e. $\gl \in \bR$, where $M(\gl)= M(\gl + i0)$.
\end{itemize}
\end{thm}
In the next corollary we find a slightly more convenient representation of the scattering matrix $\{S_\gT(\gl)\}_{\gl
\in \bR}$ of the scattering system $\{A_\gT,A_0\}$ for the case that $\Theta$ is a selfadjoint relation which is decomposed
in the form $\Theta=\Theta_\op\oplus\Theta_\infty$ with respect to $\cH=\cH_\op\oplus\cH_\infty$, cf. \eqref{t.2.14} and \eqref{t.2.15}.
If $M(\cdot)$ is the Weyl function corresponding to the boundary triplet  $\Pi= \{\kH,\Gamma_0,\Gamma_1\}$, then the function 
\begin{equation*}
 \lambda\mapsto M_\op(\lambda)=P_\op M(\lambda)\iota_\op
\end{equation*}
from \eqref{mop}
is a $[\cH_\op]$-valued Nevanlinna function, and the subspaces 
\begin{equation*}
\cH_{M_\op(\lambda)}:=\ran\bigl(\imag(M_\op(\lambda+i0))\bigr)
\end{equation*} 
of  $\cH_{M(\lambda)}$ are defined as in \eqref{2.7}.

\bc\la{t.II.2}
Let the assumptions be as in Theorem \ref{scattering}, let $\cH_{M_\op(\lambda)}$ be as above and $\kH^\infty_{M(\gl)} := \kH_{M(\gl)} \ominus \kH_{M_{\rm op}(\gl)}$. 
Then there exists a
family $V(\gl): \kH_{M(\gl)} \rightarrow \kH_{M(\gl)}$ of unitary operators
such that the representation 
\begin{equation}\la{t.2.7}
S_\gT(\gl) = V(\gl)\left\{I_{\kH^\infty_{M(\gl)}} \oplus S_{\gT_{\rm
      op}}(\gl)\right\}V(\gl)^*
\end{equation}
holds with
\begin{equation}
S_{\gT_{\rm op}}(\gl)  = I_{\kH_{M_{\rm op}(\gl)}} +
2i\sqrt{\imag(M_{\rm op}(\gl))}
\bigl(\gT_{\rm op} - M_{\rm op}(\gl)\bigr)^{-1}\sqrt{\imag(M_{\rm op}(\gl))}
\end{equation}
for a.e. $\gl \in \bR$.
\ec
\begin{proof}
Using \eqref{scatformula} and \eqref{t.2.5} we find the representation
\begin{equation}
S_\gT(\gl) = I_{\kH_{M(\gl)}} + 2i\sqrt{\imag(M(\gl))}\,\iota_\op\bigl(\gT_{\rm op} -
M_{\rm op}(\gl)\bigr)^{-1}P_{\rm op}\sqrt{\imag(M(\gl))}
\end{equation}
for a.e. $\gl \in \bR$. From the polar decomposition of $\sqrt{\imag(M(\gl))}\,\iota_\op$ we obtain a family
of isometric mappings $V_\op(\lambda)$
from $\cH_{M_\op(\lambda)}$ onto 
\begin{equation*}
\ran\bigl(\sqrt{\imag (M(\gl))}\,\iota_\op\bigr)\subset\cH_{M(\lambda)} 
\end{equation*}
defined by
\begin{equation*}
V_\op(\lambda)\sqrt{\imag(M_\op(\lambda))}:=\sqrt{\imag (M(\gl))}\,\iota_\op.
\end{equation*}
 Hence we find 
\begin{equation*}
\begin{split}
S_\gT(\gl)= I_{\kH_{M(\gl)}} + &\,\, 2i V_{\rm op}(\gl)\sqrt{\imag(M_{\rm op}(\gl))}  \\ 
&\times \bigl(\gT_{\rm op} -
M_{\rm op}(\gl)\bigr)^{-1}\sqrt{\imag(M_{\rm op}(\gl))}V_{\rm op}(\gl)^*
\end{split}
\end{equation*}
for a.e. $\gl \in \bR$. Since the Hilbert space
$\kH_{M(\gl)}$ is finite dimensional there is an isometry $V_\infty(\gl)$
acting from $\kH^\infty_{M(\gl)}= \kH_{M(\gl)} \ominus \kH_{M_{\rm op}(\gl)}$ into $\kH_{M(\gl)}$ such that
$V(\gl) := V_\infty(\gl) \oplus V_{\rm op}(\gl)$
defines a unitary operator on $\kH_{M(\gl)}$. This immediately yields \eqref{t.2.7}.
\end{proof}

\subsection{Birman-Krein formula}\label{ssfunction}

An important relation between the spectral shift function and the scattering
matrix for a pair of selfadjoint operators for the case of a trace class perturbation was found in  \cite{BK1}
by Birman and Krein. Subsequently, this relation was called the Birman-Krein formula.
Under the assumption that $A_\Theta$ and $A_0$ are selfadjoint extensions of a densely defined symmetric operator $A$ with finite deficiency 
indices and $A_\Theta$ corresponds to a selfadjoint matrix $\Theta$ via a boundary triplet $\Pi=\{\cH,\Gamma_0,\Gamma_1\}$ for $A^*$
a simple proof for the Birman-Krein formula 
\begin{equation*}
\det(S_{\gT}(\gl)) = \exp\bigl(-2\pi i \xi_\gT(\gl)\bigr)
\end{equation*}
was given in \cite{BeMN1}. Here $\xi_\gT(\cdot)$ is the spectral shift function of the pair $\{A_\Theta,A_0\}$  defined by \eqref{t.2.13} 
and the scattering matrix $\{S_\gT(\gl)\}_{\gl \in \bR}$ is given by \eqref{scatformula}. 

The following theorem generalizes \cite[Theorem 4.1]{BeMN1} to the case of a selfadjoint relation $\Theta$ (instead of a matrix), so that the Birman-Krein formula 
is verified for all pairs of selfadjoint extensions of the underlying symmetric operator.
\begin{thm}\label{V.1a}
Let $A$ be a densely defined closed simple symmetric operator in the separable Hilbert space
$\gotH$ with equal finite deficiency indices, let $\Pi=
\{\kH,\Gamma_0,\Gamma_1\}$ be a boundary triplet for $A^*$ and let
$M(\cdot)$ be the corresponding Weyl function. Furthermore, let
$A_0=A^*\!\upharpoonright\ker(\Gamma_0)$  and let
$A_\gT=A^*\upharpoonright\Gamma^{-1}\gT$, $\Theta\in\widetilde\cC(\cH)$,
be a selfadjoint extension of $A$ in $\sH$. Then the spectral shift function $\xi_\gT(\cdot)$
in \eqref{xitheta} and the scattering matrix 
$\{S_\gT(\lambda)\}_{\lambda\in\bR}$ of the pair 
$\{A_\gT,A_0\}$  are  related via
\begin{equation}\label{5.6a}
\det\bigl(S_{\gT}(\gl)\bigr) = \exp\bigl(-2\pi i \xi_\gT(\gl)\bigr)
\end{equation}
for a.e. $\lambda\in\dR$. 
\end{thm}
\begin{proof}
To verify the Birman-Krein formula we note that by \eqref{det}  
\begin{equation*}
\begin{split}
&\exp\bigl(-2i\imag\bigl(\tr(\log(M_{\rm op}(\gl) - \gT_{\rm op}))\bigr) \bigr) \\
&\qquad=\exp\bigl(-\tr(\log(M_{\rm op}(\gl) - \gT_{\rm op}))\bigr)
\exp\bigl(\,\overline{\tr(\log(M_{\rm op}(\gl) - \gT_{\rm op}))}\,\bigr)\\
&\qquad=\frac{\overline{\det(M_{\rm op}(\gl) - \gT_{\rm op})}}{\det(M_{\rm op}(\gl) - \gT_{\rm op})}
= \frac{\det(M_{\rm op}(\gl)^* - \gT_{\rm op})}{\det(M_{\rm op}(\gl) - \gT_{\rm op})}
\end{split}
\end{equation*}
holds for all $\lambda\in\dC_+$. Hence we find
\begin{equation}\label{5.13}
\exp\bigl(-2\pi i\xi_\gT(\lambda)\bigr)= 
\frac{\det\bigl(M_{\rm op}(\gl + i0)^* - \gT_{\rm op}\bigr)}
{\det\bigl(M_{\rm op}(\gl + i0) - \gT_{\rm op}\bigr)}
\end{equation}
for a.e. $\gl \in \bR$,  where $M_{\rm op}(\gl + i0) := 
\lim_{\epsilon\to+0}M_{\rm op}(\gl + i\epsilon)$ 
exists for a.e. $\gl \in \bR$. It follows from the representation
of the scattering matrix in Corollary~\ref{t.II.2} and the identity $\det(I+A B)=\det(I+B A)$ that
\begin{equation*}\begin{split}
&\det S_\gT(\gl)\\ 
& \quad
=\det\left(I_{\kH_\op}+2i\bigl(\imag (M_{\rm op}(\lambda+i0))\bigr)
\bigl(\gT_{\rm op}-M_{\rm op}(\gl+i 0)\bigr)^{-1} \right) \\
& \quad
=\det\left(I_{\kH_\op}+\bigl(M_{\rm op}(\gl+i 0)-M_{\rm op}(\gl+i 0)^*\bigr)
\bigl(\gT_{\rm op}-M_{\rm op}(\gl+i 0)\bigr)^{-1}\right)   \\
& \quad
=\det\left(\bigl(M_{\rm op}(\gl+i 0)^*-\Theta_\op\bigr)\cdot
\bigl(M_{\rm op}(\gl+i 0)-\Theta_\op\bigr)^{-1}\right) \\
& \quad
=\frac{\det\bigl(M_{\rm op}(\gl+i 0)^*-\Theta_\op\bigr)}
{\det\bigl(M_{\rm op}(\gl+i 0)-\Theta_\op\bigr)}
\end{split}
\end{equation*}
holds for a.e. $\gl \in \bR$. 
Comparing this with \eqref{5.13} we obtain \eqref{5.6a}. 
\end{proof}

\section{Dissipative scattering systems}\la{dilations}

In this section we investigate scattering systems consisting of a
maximal dissipative and a selfadjoint operator, which are both extensions of a common symmetric operator with equal finite deficiency indices.
We shall explicitly construct a so-called dilation of the maximal dissipative operator and we calculate the spectral shift function
of the dissipative scattering system with the help of this dilation. It will be shown that the scattering matrix of the dissipative scattering system
and this spectral shift function are connected via a modified Birman-Krein formula.

\subsection{Selfadjoint dilations of maximal dissipative operators}\label{sec3.1}

Let $A$ be a densely defined closed
simple symmetric operator in the separable Hilbert space
$\gotH$ with equal finite deficiency indices $n_+(A)=n_-(A)=n<\infty$,
let $\gP = \{\kH,\gG_0,\gG_1\}$,
$A_0=A^*\upharpoonright\ker(\Gamma_0)$, 
be a boundary triplet for $A^*$
and let $D \in [\kH]$ be a dissipative $n\times n$-matrix, i.e. $\imag (D)\leq 0$.
Then by Proposition~\ref{propo}~(iii) the closed extension
\begin{equation*}
A_D=A^*\upharpoonright\ker(\Gamma_1-D\Gamma_0)
\end{equation*}
of $A$ corresponding to $\gT = D$ via \eqref{bij} is maximal
dissipative, that is, $A_D$ is dissipative and maximal in the sense that each dissipative extension of $A_D$ in $\sH$ coincides with $A_D$.
Observe that $\bC_+$ belongs to $\rho(A_D)$. 
For $\gl \in \rho(A_D)\cap\rho(A_0)$ the resolvent of
the extension $A_D$ is given by
\begin{equation}\la{3.1}
(A_D -\gl)^{-1} = (A_0 - \gl)^{-1} + \gga(\gl)\bigl(D -
M(\gl)\bigr)^{-1}\gga(\bar{\gl})^*,
\end{equation}
cf. \eqref{resoll}.
With respect to the decomposition
\begin{equation*}
D=\real (D)+i\imag (D)
\end{equation*}
we decompose $\cH$ into the orthogonal sum of the finite dimensional
subspaces $\ker(\imag(D))$ and $\cH_D:=\ran(\imag(D))$,
\begin{equation}\label{decoh}
\cH=\ker(\imag(D))\oplus\cH_D,
\end{equation}
and denote by $P_D$ the orthogonal projection from $\cH$ onto $\cH_D$ and by $\iota_D$ the canonical embedding of $\cH_D$ into $\cH$. 
Since $\imag(D)\leq 0$ the selfadjoint matrix
\begin{equation*}
-P_D\imag(D)\,\iota_D \in[\cH_D]
\end{equation*}
is strictly positive and therefore (see, e.g. \cite{DHMS06,DM95})  the function 
\begin{equation}\label{taud1}
\lambda\mapsto \begin{cases} -iP_D\imag(D)\iota_D, & \lambda\in\dC_+,\\
i P_D\imag(D)\,\iota_D, & \lambda\in\dC_-,\end{cases}
\end{equation}
can be realized as the Weyl function corresponding to a boundary triplet of a symmetric operator. 

Here the symmetric operator and boundary triplet can be made more explicit, cf. \cite[Lemma 3.1]{BeMN2}. In fact, let  
$G$ be the symmetric first order differential operator in the Hilbert
space $L^2(\dR,\cH_D)$ defined by
\begin{equation}\la{t.3.3}
(Gg)(x)=-ig^\prime(x),\qquad\dom (G)=\bigl\{g\in W^1_2(\dR,\cH_D)\,:\,g(0)=0\bigr\}.
\end{equation}
Then 
$G$ is simple, $n_\pm(G)=\dim\cH_D$ and the adjoint operator
$G^*g=-ig^\prime$ is defined on 
\begin{equation*}
\dom (G^*)=W^1_2(\dR_-,\cH_D)\oplus W^1_2(\dR_+,\cH_D).
\end{equation*}
Moreover, the triplet $\Pi_G=\{\cH_D,\gY_0,\gY_1\}$,
where
\begin{equation}\la{t.3.4}
\begin{split}
\gY_0 g&:=\frac{1}{\sqrt{2}}
\bigl(-P_D\imag(D)\,\iota_D\bigr)^{-\frac{1}{2}}\bigl(g(0+)-g(0-)\bigr),\\
\gY_1 g&:=\frac{i}{\sqrt{2}}\bigl(-P_D\imag(D)\,\iota_D\bigr)^{\frac{1}{2}}
\bigl(g(0+)+g(0-)\bigr),
\end{split}
\end{equation}
$g\in\dom(G^*)$, is a boundary triplet for 
$G^*$ and the extension $G_0:=G^*\upharpoonright\ker(\gY_0)$ of $G$ is the usual
selfadjoint first order differential operator in 
$L^2(\dR,\cH_D)$ with domain $\dom(G_0)=W^{1}_2(\dR,\cH_D)$
and $\sigma(G_0)=\dR$.
It is not difficult to see that the defect subspaces of $G$ are given by
\begin{equation*}
\ker(G^*-\lambda)=\begin{cases}\sp \{x\mapsto e^{i\lambda x}\chi_{\dR_+}(x)\xi\,:\,\xi\in\cH_D\}, & \lambda\in\dC_+,\\
                   \sp \{x\mapsto e^{i\lambda x}\chi_{\dR_-}(x)\xi\,:\,\xi\in\cH_D\}, & \lambda\in\dC_-,
                  \end{cases}
\end{equation*}
and therefore it follows that  the Weyl function $\tau(\cdot)$ corresponding to 
the boundary triplet $\Pi_G=\{\cH_D,\gY_0,\gY_1\}$ is
given by
\begin{equation}\label{taud}
\tau(\lambda)=\begin{cases} -iP_D\imag(D)\,\iota_D, & \lambda\in\dC_+,\\
i P_D\imag(D)\,\iota_D, & \lambda\in\dC_-.\end{cases}
\end{equation}

Let $A$ be the densely defined closed simple symmetric operator in $\gotH$ from above and let $G$
be the first order differential operator in \eqref{t.3.3}.
Clearly
\begin{equation*}
K := \begin{pmatrix} A & 0 \\ 0 & G\end{pmatrix}
\end{equation*}
 is a densely defined closed simple symmetric operator in the separable
Hilbert space
\begin{equation*}
\gotK:=\gotH\oplus L^2(\dR,\cH_D)
\end{equation*}
with equal finite deficiency indices $n_\pm(K)=n_\pm(A)+n_\pm(G)=n+\dim\cH_D<\infty$ and the adjoint
is 
\begin{equation*}
K^*=\begin{pmatrix} A^* & 0 \\ 0 &  G^*\end{pmatrix}.
\end{equation*}
The elements in $\dom(K^*)=\dom(A^*)\oplus\dom(G^*)$ will be
written in the form $f\oplus g$, $f\in\dom(A^*)$, $g\in\dom
(G^*)$. It is straightforward to check that
$\widetilde{\gP}= \{\widetilde{\kH},\widetilde{\gG}_0,\widetilde{\gG}_1\}$, where $\widetilde{\kH} := \kH \oplus \kH_D$,
\begin{equation}\la{3.12}
 \widetilde{\gG}_0(f\oplus g) :=
\begin{pmatrix}\gG_0 f \\ \gY_0 g\end{pmatrix}\quad\text{and} \quad \widetilde{\gG}_1(f\oplus g) :=
\begin{pmatrix}\gG_1f-\real(D)\Gamma_0f\\ \gY_1g\end{pmatrix},
\end{equation}
$f\oplus g\in\dom(K^*)$, is a boundary triplet for $K^*$.
If $\gamma(\cdot),\nu(\cdot)$ and $M(\cdot),\tau(\cdot)$ are the $\gamma$-fields and Weyl functions
of the boundary triplets $\Pi=\{\cH,\Gamma_0,\Gamma_1\}$ and $\Pi_G=\{\cH_D,\gY_0,\gY_1\}$,
respectively, then one easily verifies that the Weyl function
$\widetilde M(\cdot)$ and $\gamma$-field $\widetilde\gamma(\cdot)$ corresponding to the boundary triplet $\widetilde{\gP}= \{\widetilde{\kH},\widetilde{\gG}_0,\widetilde{\gG}_1\}$ are
given by
\begin{equation}\la{3.14a}
\widetilde{M}(\gl) =
\begin{pmatrix}
M(\gl)-\real(D) & 0\\
0 & \gt(\gl)
\end{pmatrix}, \qquad \gl \in \dC\backslash\dR,
\end{equation}
and
\begin{equation}\label{3.14aa}
\widetilde{\gga}(\gl) =
\begin{pmatrix}
\gga(\gl) & 0\\
0 & \nu(\gl)
\end{pmatrix} , \qquad \gl \in \dC\backslash\dR,
\end{equation}
respectively. Observe that 
\begin{equation}\label{k0}
K_0:=K^*\upharpoonright\ker(\widetilde\Gamma_0)=\begin{pmatrix}A_0 & 0 \\ 0 & G_0\end{pmatrix}
\end{equation}
holds. With respect to the decomposition 
\begin{equation*}
\widetilde{\kH} = \ker(\imag(D)) \oplus \kH_D \oplus
\kH_D
\end{equation*}
 of $\widetilde\cH$ (cf. \eqref{decoh}) we define the linear relation $\widetilde{\gT}$ in $\widetilde\cH$ by
\begin{equation}\label{wtdef}
\widetilde\gT:=\left\{\begin{pmatrix}(u,v,v)^\top\\ (0,-w,w)^\top\end{pmatrix}:
u\in\ker(\imag(D),\; v,w\in\cH_D\right\}.
\end{equation}
We leave it to the reader to check that $\widetilde\gT$ is selfadjoint.
Hence by Proposition~\ref{propo} the operator 
\begin{equation*}
\begin{split}
\widetilde K:&=K_{\widetilde\gT}=
K^*\upharpoonright\widetilde\Gamma^{-1}\widetilde\gT\\
&=\left\{f\oplus g\in\dom (A^*)\oplus\dom (G^*):\bigl (\widetilde\Gamma_0(f\oplus g), \widetilde\Gamma_1(f\oplus g)\bigr)^\top\in\widetilde\Theta\right\}
\end{split}
\end{equation*}
is a selfadjoint extension of the symmetric 
operator $K$ in $\gotK=\gotH\oplus L^2(\dR,\cH_D)$. 
The following theorem was proved in \cite{BeMN2}, see also \cite{P1,P2} for a special case involving Sturm-Liouville
operators with dissipative boundary conditions.
\begin{thm}\label{III.1}
Let $A$, $\Pi=\{\cH,\Gamma_0,\Gamma_1\}$ and $A_D=A^*\!\upharpoonright\!\ker(\Gamma_1-D\Gamma_0)$ be as above.
Furthermore, 
let $G$ and $\Pi_G=\{\cH_D,\gY_0,\gY_1\}$
be given by \eqref{t.3.3} and \eqref{t.3.4}, respectively, and let $K=A\oplus G$. Then the
selfadjoint extension $\widetilde{K} $ of $K$ has the form
\begin{equation}\label{widetildek}
\widetilde K=K^*\upharpoonright\left\{f\oplus g\in\dom(K^*):
\begin{matrix}
P_D\Gamma_0f-\gY_0 g=0,\\(I-P_D)(\Gamma_1-\real(D)\Gamma_0)f=0,\\
P_D(\Gamma_1-\real(D)\Gamma_0)f+\gY_1 g=0
\end{matrix}\right\}
\end{equation}
and $\widetilde K$ is a minimal selfadjoint dilation of the maximal dissipative
operator $A_D$, that is, for all $\lambda\in\dC_+$
\begin{equation*}
P_\gotH\bigl(\widetilde K-\lambda\bigr)^{-1}\upharpoonright_{\gotH}=(A_D-\lambda)^{-1}
\end{equation*}
holds and the minimality condition
$\gotK=\clospa\{(\widetilde K-\lambda)^{-1}\gotH:\lambda\in\dC\backslash\dR\}$
is satisfied. Moreover $\sigma(\widetilde K)=\dR$.
\end{thm}

We note that also in the case where the parameter $D$ is not a dissipative matrix
but a maximal dissipative relation in $\cH$ a minimal selfadjoint dilation of $A_D$ can
be constructed in a similar way as in Theorem~\ref{III.1}, see \cite[Remark 3.3]{BeMN2}

\subsection{Spectral shift function and trace formula}

In order to calculate the spectral shift function of the pair $\{\widetilde K,K_0\}$ from \eqref{k0} and \eqref{widetildek} we write the selfadjoint relation $\widetilde\Theta$ from \eqref{wtdef}
in the form 
$\widetilde{\gT} = \widetilde{\gT}_{\rm op} \oplus
\widetilde{\gT}_\infty$, where
\begin{equation}\label{wtdefop}
\widetilde{\gT}_{\rm op} :=
\left\{
\begin{pmatrix}
(u,v,v)^\top\\
(0,0,0)^\top
\end{pmatrix}:
\; u \in \ker(\imag(D)),\; v \in \cH_D
\right\}
\end{equation}
is the zero operator in the space 
\begin{equation}
\widetilde{\kH}_\op := 
\left\{
\begin{pmatrix} 
u\\
v\\
v
\end{pmatrix}: 
\;u \in \ker(\imag(D)),\;v \in \cH_D
\right\}
\end{equation}
and
\begin{equation}
\widetilde{\gT}_\infty :=
\left\{
\begin{pmatrix}
(0,0,0)^\top\\
(0,-w,w)^\top
\end{pmatrix}:
\; w\in \cH_D
\right\}
\end{equation}
is the purely multivalued relation in the space 
\begin{equation}
\widetilde\kH_\infty = \widetilde{\kH} \ominus
\widetilde\kH_\op=
\left\{
\begin{pmatrix}
0\\
-w\\
w
\end{pmatrix}
: \; w \in \cH_D\right\}.
\end{equation}
The orthogonal projection from $\widetilde\cH$ onto $\widetilde\cH_\op$ will be denoted by $\widetilde P_\op$ and the canonical embedding of $\widetilde\cH_\op$ in $\widetilde\cH$
is denoted by $\widetilde\iota_\op$. 
As an immediate consequence of Theorem~\ref{V.1} we find the following representation of a spectral shift function for the pair $\{\widetilde K,K_0\}$.

\begin{cor}
Let $A$ and $G$ be the symmetric operators from Section~\ref{sec3.1} and let $K = A \oplus G$. Furthermore, let $\widetilde{\gP}
= \{\widetilde{\kH},\widetilde{\gG}_0,\widetilde{\gG}_1\}$ 
be the boundary triplet for $K^*$ from \eqref{3.12} with Weyl function $\widetilde{M}(\cdot)$ given
by \eqref{3.14a} and define the $[\widetilde\cH_\op]$-valued Nevanlinna function by
\begin{equation*}
\widetilde M_\op(\lambda):=\widetilde P_\op \widetilde M(\lambda)\,\widetilde\iota_\op.
\end{equation*}
Then the limit $\lim_{\epsilon\rightarrow +0}\widetilde M_\op(\lambda+i\epsilon)$ exists for a.e. $\lambda\in\dR$ and the function
\begin{equation}\label{xiwt}
\xi_{\widetilde{\gT}}(\gl):=\frac{1}{\pi}\imag\bigl(\tr\bigl(\log(\widetilde{M}_{\rm op}(\gl+i0)\bigr)\bigr)
\end{equation}
is a spectral shift function for the pair $\{\widetilde K,K_0\}$ with $0\leq\xi_{\widetilde{\gT}}(\gl)\leq\dim\widetilde\cH_\op=n$.
\end{cor}

Observe that the spectral shift function in \eqref{xiwt} satisfies
the trace formula
\begin{equation}\la{t.4.15}
\tr\bigl((\widetilde{K} - \gl)^{-1} - (K_0 - \gl)^{-1}\bigr) =
-\int_\bR \frac{1}{(t - \gl)^2}\,\xi_{\widetilde{\gT}}(t)\;dt
\end{equation}
for $\gl \in \bC \backslash \bR$. In the following theorem we calculate the spectral shift function of $\{\widetilde{K},K_0\}$ in a more explicit form up to a constant $2k$, $k\in\dZ$.
We mention that the spectral shift function in \eqref{t.4.16} below can be regarded as the spectral shift function of the dissipative scattering system $\{A_D,A_0\}$, cf. \cite{N84,N85,N86}.
\bt\la{III.8}
Let $A$ and $G$ be the symmetric operators from Section~\ref{sec3.1} and let 
$\gP = \{\kH,\gG_0,\gG_1\}$ be a boundary triplet for $A^*$ with corresponding Weyl function $M(\cdot)$. Let $D\in[\cH]$ be a dissipative $n\times n$-matrix 
and let $A_D=A^*\upharpoonright\ker(\Gamma_1-D\Gamma_0)$ be the corresponding maximal dissipative extension of $A$. Furthermore, let $K_0$ be as in \eqref{k0} and let 
$\widetilde K$ be the minimal selfadjoint dilation of $A_D$ from \eqref{widetildek}.

Then the spectral
shift function $\xi_{\widetilde{\gT}}(\cdot)$ of the pair $\{\widetilde{K},K_0\}$ admits the
representation $\xi_{\widetilde{\gT}}(\cdot) = \eta_D(\cdot)+2k$ for some $k\in\dZ$, where
\begin{equation}\la{t.4.16}
\eta_D(\gl) := 
\frac{1}{\pi}\imag\bigl(\tr\bigl(\log(M(\gl+i0) - D)\bigr)\bigr)
\end{equation}
for a.e. $\gl \in \bR$, and
the modified trace formulas 
\begin{equation}\la{t.4.17}
\tr\bigl((A_D - \gl)^{-1} - (A_0 - \gl)^{-1}\bigr) = 
-\int_\bR  \;\frac{1}{(t - \gl)^2}\,\eta_D(t)\;dt, \quad \gl \in
\bC_+,
\end{equation}
and
\begin{equation}\la{t.4.18}
\tr\bigl((A^*_D - \gl)^{-1} - (A_0 - \gl)^{-1}\bigr) = 
-\int_\bR  \;\frac{1}{(t - \gl)^2}\,\eta_D(t)\;dt, \quad \gl \in
\bC_-,
\end{equation}
are valid. 
\et
\begin{proof}
With the help of the operator 
\begin{equation*}
V:\kH \longrightarrow \kH,\quad x\mapsto \begin{pmatrix} (I-P_D)x \\ \tfrac{1}{\sqrt{2}}P_D x \end{pmatrix}
\end{equation*}
and the unitary operator
\begin{equation*}
 \widetilde V:\kH \longrightarrow
\widetilde\kH_\op,\quad x\mapsto \begin{pmatrix} (I-P_D)x \\ \tfrac{1}{\sqrt{2}}P_D x\\ \tfrac{1}{\sqrt{2}}P_D x\end{pmatrix}
\end{equation*}
one easily verifies that
\begin{equation}\la{t.4.21}
\begin{split}
\widetilde V^*\widetilde{M}_{\rm op}(\gl)\widetilde V&  = V 
\left(M(\gl) - \real(D) + \begin{pmatrix} 0 & 0 \\ 0 &\gt(\gl)\end{pmatrix}\right)V\\ 
&=V(M(\gl) - D)V
\end{split}
\end{equation}
holds for all $\lambda\in\dC_+$.
Using this relation and the definition of $\log(\cdot)$ in \eqref{log} we get
\begin{equation*}
\tr\bigl(\log\bigl(\widetilde{M}_{\rm op}(\gl)\bigr)\bigr) = 
\tr\bigl(\log\bigl(\widetilde V^*\widetilde{M}_{\rm op}(\gl)\widetilde V\bigr)\bigr)=\tr\bigl(\log\bigl ( V(M(\gl) - D)V \bigr)\bigr)
\end{equation*}
and therefore  \eqref{detlog} (see also \cite{GK}) implies 
\begin{equation*}
\begin{split}
&\frac{d}{d\lambda}\tr\bigl(\log\bigl(\widetilde{M}_{\rm op}(\gl)\bigr)\bigr)
=\frac{d}{d\lambda} \log\bigl(\det\bigl( V(M(\gl) - D)V \bigr) \bigr)\\
&\quad=\frac{d}{d\lambda} \log\bigl(\det (M(\gl) - D)\bigr)+\frac{d}{d\lambda} \log\bigl(\det V^2\bigr)
=\frac{d}{d\lambda}\tr\bigl(\log(M(\gl) - D)\bigr).
\end{split}
\end{equation*}
Hence $\tr(\log(\widetilde{M}_{\rm op}(\cdot)))$ and $\tr(\log(M(\cdot) - D))$ differ by a constant.
From 
\begin{equation*}
\exp\bigl(\tr\bigl(\log\bigl(\widetilde{M}_{\rm op}(\gl)\bigr)\bigr)\bigr)
= \exp\bigl( \tr\bigl(\log(M(\lambda)-D)\bigr) \bigr)\det V^2
\end{equation*}
we conclude that there exists $k\in\dZ$ such that
\begin{equation*}
\imag\bigl(\tr\bigl(\log\bigl(\widetilde{M}_{\rm op}(\gl)\bigr)\bigr)\bigr)= \imag\bigl(\tr\bigl(\log(M(\lambda)-D)\bigr)\bigr)+2k\pi  
\end{equation*}
holds. Hence it follows that the spectral shift function $\xi_{\widetilde{\gT}}$ of the pair $\{\widetilde K,K_0\}$ in \eqref{xiwt} and the function $\eta_D(\cdot)$ in \eqref{t.4.16} differ by 
$2k$ for some $k\in\dZ$.

Next we verify that the trace formulas \eqref{t.4.17} and \eqref{t.4.18} hold. From \eqref{resoll} we obtain
\begin{equation}
\begin{split}
\tr\bigl((\widetilde K - \gl)^{-1} - (K_0 - \gl)^{-1}\bigr)
&= \tr\bigl(\widetilde{\gga}(\gl)
\bigl(\widetilde{\gT} -  \widetilde{M}(\gl)\bigr)^{-1}\widetilde{\gga}(\bar{\gl})^*\bigr)\\
&=
\tr\bigl(\bigl(\widetilde{\gT} -\widetilde{M}(\gl)\bigr)^{-1}
\widetilde{\gga}(\bar{\gl})^*\widetilde{\gga}(\gl)\bigr)
\end{split}
\end{equation}
for $\lambda\in\dC\backslash\dR$.
As in \eqref{5.9} and \eqref{5.10} we find
\begin{equation}
\tr\bigl((\widetilde K - \gl)^{-1} - (K_0 - \gl)^{-1}\bigr) =
\tr\left(
\bigl(\widetilde{\gT} -\widetilde{M}(\gl)\bigr)^{-1}
\frac{d}{d\gl}\widetilde{M}(\gl)
\right).
\end{equation}
With the same argument as in the proof of Theorem~\ref{V.1} we then conclude
\begin{equation}\label{qwert}
\tr\bigl((\widetilde K - \gl)^{-1} - (K_0 - \gl)^{-1}\bigr) =
\tr\left(
\bigl(\widetilde{\gT}_{\rm op} -\widetilde{M}_{\rm op}(\gl)\bigr)^{-1}
\frac{d}{d\gl}\widetilde{M}_{\rm op}(\gl)
\right).
\end{equation}
Since $\widetilde{\gT}_{\rm op} = 0$ and $\widetilde V$ is unitary it follows from \eqref{t.4.21}
that
\begin{equation*}
\bigl(\widetilde{\gT}_{\rm op} -\widetilde{M}_{\rm op}(\gl)\bigr)^{-1} =-\widetilde{M}_{\rm op}(\gl)^{-1}=-\widetilde V V^{-1}\bigl(M(\lambda)-D\bigr)^{-1}V^{-1}\widetilde V^*
\end{equation*}
and
\begin{equation}
\frac{d}{d\gl}\widetilde{M}_{\rm op}(\gl)=\widetilde V V \frac{d}{d\lambda} M(\lambda)V\widetilde V^* 
\end{equation}
holds. This together with \eqref{qwert} implies
\begin{equation*}
\tr\bigl((\widetilde K - \gl)^{-1} - (K_0 - \gl)^{-1}\bigr) =
\tr\left(-
(M(\lambda)-D)^{-1}
\frac{d}{d\gl} M(\gl)
\right)
\end{equation*}
for all $\lambda\in\dC_+$ and with \eqref{5.9} we get
\begin{equation*}
\tr\bigl((\widetilde K - \gl)^{-1} - (K_0 - \gl)^{-1}\bigr) =
\tr\left(\gamma(\lambda)
(D-M(\lambda))^{-1}\gamma(\bar\lambda)^*
\right)
\end{equation*}
as in \eqref{5.10}.  Using \eqref{3.1} we obtain
\begin{equation}
\tr\bigl((\widetilde K - \gl)^{-1} - (K_0 - \gl)^{-1}\bigr ) =
\tr\left((A_D - \gl)^{-1} - (A_0 - \gl)^{-1}\right)
\end{equation}
for $\gl \in \bC_+$. Taking into account \eqref{t.4.15} we prove
\eqref{t.4.17} and \eqref{t.4.18} follows by taking adjoints. 
\end{proof}

\subsection{Scattering matrices of dissipative and Lax-Phillips scattering systems}

In this section we recall some results from \cite{BeMN2} on the interpretation of the diagonal entries of the scattering matrix of $\{\widetilde K,K_0\}$ 
as scattering matrices of a dissipative and a Lax-Phillips scattering system.
For this, let again $A$ and $G$ be the symmetric operators from Section~\ref{sec3.1} and let 
$\gP = \{\kH,\gG_0,\gG_1\}$ be a boundary triplet for $A^*$ with Weyl function $M(\cdot)$. Let $D\in[\cH]$ be a dissipative $n\times n$-matrix 
and let $A_D=A^*\!\upharpoonright\!\ker(\Gamma_1-D\Gamma_0)$ be the corresponding maximal dissipative extension of $A$. 
Furthermore, let $\widetilde{\gP}
= \{\widetilde{\kH},\widetilde{\gG}_0,\widetilde{\gG}_1\}$  be  the boundary triplet for $K^*=A^*\oplus G^*$ from \eqref{3.12} with Weyl function $\widetilde{M}(\cdot)$ given
by \eqref{3.14a}, let $\widetilde\Theta$ be as in \eqref{wtdef} and let $\widetilde K$ be the minimal selfadjoint dilation of $A_D$ given by \eqref{widetildek}.
It follows immediately from Theorem \ref{scattering} that the
scattering matrix $\{\widetilde S(\gl)\}_{\gl \in \bR}$ of the complete
scattering system $\{\widetilde{K},K_0\}$ is given by 
\begin{equation*}
\widetilde{S}(\gl) = I_{\widetilde{\kH}_{\widetilde{M}(\gl)}} +
2i\sqrt{\imag(\widetilde{M}(\gl))}\bigl(\widetilde{\gT} -
\widetilde{M}(\gl)\bigr)^{-1}
\sqrt{\imag(\widetilde{M}(\gl))}
\end{equation*}
in the spectral representation $L^2(\bR,d\gl,\kH_{\widetilde{M}(\gl)})$ of $K_0^{ac}$. Here the spaces 
\begin{equation*}
\widetilde{\kH}_{\widetilde{M}(\gl)}:=\ran\bigl(\imag(\widetilde M(\lambda+i0))\bigr)
\end{equation*}
for a.e. $\lambda\in\dR$ are defined in analogy to \eqref{2.7}. This representation can be made more explicit, cf. \cite[Theorem 3.6]{BeMN2}.

\begin{thm}\label{dilscat}
Let $A$, $\Pi=\{\cH,\Gamma_0,\Gamma_1\}$, $M(\cdot)$ and $A_D$ 
be as above, let 
$K_0=A_0\oplus G_0$ and let $\widetilde K$ be the minimal
selfadjoint dilation of $A_D$ from Theorem~{\rm \ref{III.1}}. Then
the following holds:
\begin{itemize}
\item[{\rm (i)}] $K^{ac}_0=A^{ac}_0\oplus G_0$ is unitarily 
equivalent to the multiplication operator
with the free variable in $L^2(\bR,d\gl,\kH_{M(\gl)}\oplus\cH_D)$.

\item[{\rm (ii)}] In $L^2(\bR,d\gl,\kH_{M(\gl)}\oplus\cH_D)$
the scattering matrix $\{\widetilde S(\gl)\}_{\lambda\in\dR}$ of the 
complete scattering system $\{\widetilde K,K_0\}$ is given by

\begin{displaymath}
\widetilde{S}(\gl) = 
\begin{pmatrix}
I_{\kH_{M(\gl)}} & 0 \\ 
0 & I_{\kH_D} 
\end{pmatrix} + 2i
\begin{pmatrix}
\widetilde T_{11}(\lambda) &  \widetilde T_{12}(\lambda)\\
 \widetilde T_{21}(\lambda) &  \widetilde T_{22}(\lambda)
\end{pmatrix}
\in[\cH_{M(\lambda)}\oplus\cH_D],
\end{displaymath}
for a.e. $\lambda\in\dR$, where
% %
\begin{equation*}
\begin{split}
\widetilde T_{11}(\lambda)&=\sqrt{\imag(M(\lambda))}
\bigl(D - M(\gl)\bigr)^{-1}\sqrt{\imag(M(\lambda))},\\
\widetilde T_{12}(\lambda)&=\sqrt{\imag(M(\lambda))}
                            \bigl(D - M(\gl)\bigr)^{-1}\sqrt{-\imag(D)},\\
\widetilde T_{21}(\lambda)&=\sqrt{-\imag(D)}
\bigl(D - M(\gl)\bigr)^{-1}\sqrt{\imag(M(\lambda))},\\
\widetilde T_{22}(\lambda)&=\sqrt{-\imag(D)}\bigl(D - M(\gl)\bigr)^{-1}\sqrt{-\imag(D)}\\
\end{split}
\end{equation*}
and $M(\lambda)=M(\lambda+i0)$.
\end{itemize}
\end{thm}
Observe that the scattering matrix $\{\widetilde S(\lambda)\}_{\lambda\in\dR}$ of the scattering system
$\{\widetilde K,K_0\}$ depends only on the dissipative matrix $D$ and the
Weyl function $M(\cdot)$ of the boundary triplet $\Pi=\{\cH,\Gamma_0,\Gamma_1\}$ for $A^*$, i.e., 
$\{\widetilde S(\lambda)\}_{\lambda\in\dR}$ is completely determined
by objects corresponding to the operators $A,A_0$ and $A_D$ in $\sH$.

In the following we will focus on the so-called
{\it dissipative scattering system} $\{A_D,A_0\}$ and we refer the reader to
\cite{D3,D2,Ma1,Na1,Na2,N81,N84,N85,N86} for a detailed investigation of such scattering systems.
We recall that the wave operators $W_\pm(A_D,A_0)$ of the dissipative scattering
system $\{A_D,A_0\}$ are defined by
\begin{equation*}
W_+(A_D,A_0)=\slim_{t\rightarrow +\infty} e^{it A_D^*} e^{-itA_0}P^{ac}(A_0)
\end{equation*}
and
\begin{equation*}
W_-(A_D,A_0)=\slim_{t\rightarrow +\infty} e^{-it A_D} e^{itA_0}P^{ac}(A_0).
\end{equation*}
The scattering operator
\begin{equation*}
S_D:=W_+(A_D,A_0)^*W_-(A_D,A_0)
\end{equation*}
of the dissipative scattering system $\{A_D,A_0\}$ 
will be regarded as an operator in $\gotH^{ac}(A_0)$. Then $S_D$
is a contraction which in general is not unitary. 
Since $S_D$ and $A_0^{ac}$ commute it follows that $S_D$ is unitarily
equivalent to a multiplication operator induced by a 
family $\{S_D(\lambda)\}_{\lambda\in\dR}$ of contractive operators
in a spectral representation of $A_0^{ac}$.

With the help of Theorem~\ref{dilscat} 
we obtain a representation of the scattering matrix of
the dissipative scattering system $\{A_D,A_0\}$ in terms of the Weyl function $M(\cdot)$
of $\Pi=\{\cH,\Gamma_0,\Gamma_1\}$ in the following corollary, cf. \cite[Corollary 3.8]{BeMN2}.
\begin{cor}\label{disscatcor}
Let $A$, $\Pi=\{\cH,\Gamma_0,\Gamma_1\}$, 
$A_0=A^*\!\upharpoonright\ker(\Gamma_0)$ and $M(\cdot)$ be as above and let 
$A_D=A^*\!\upharpoonright\ker(\Gamma_1-D\Gamma_0)$, $D\in[\cH]$,  be maximal dissipative. Then the following holds:

\begin{itemize}
\item [{\rm (i)}] $A^{ac}_0$ is unitarily equivalent to the multiplication operator
with the free variable in $L^2(\bR,d\gl,\kH_{M(\gl)})$.

\item [{\rm (ii)}] The scattering matrix $\{S_D(\gl)\}$ of the dissipative scattering system $\{A_D,A_0\}$ is given by
the left upper corner of the scattering matrix $\{\widetilde S(\lambda)\}$ in Theorem~\ref{dilscat}, i.e.
\begin{displaymath}
S_D(\gl) =I_{\kH_{M(\gl)}}+2i 
\sqrt{\imag(M(\lambda))}\bigl(D - M(\gl)\bigr)^{-1}\sqrt{\imag(M(\lambda))}
\end{displaymath}
for all a.e. $\gl \in \bR$, where $M(\lambda)=M(\lambda+i0)$. 
\end{itemize}
\end{cor}
In the following we are going to interpret the right lower corner
of the scattering matrix $\{\widetilde S(\lambda)\}$ of $\{\widetilde K,K_0\}$ as the scattering matrix corresponding
to a Lax-Phillips scattering system, see e.g. \cite{BW,LP} for further details.
To this end we decompose the space
$L^2(\dR,\cH_D)$ into the orthogonal sum of the subspaces
\begin{equation}\label{dpm}
\kD_-:=L^2(\dR_-,\cH_D)\quad\text{and}\quad\kD_+:=L^2(\dR_+,\cH_D).
\end{equation}
Then clearly 
\begin{equation*}
\gotK=\gotH\oplus L^2(\dR,\cH_D) =\gotH\oplus\cD_-\oplus\cD_+
\end{equation*} 
and we agree to denote the elements
in $\gotK$ in the form $f\oplus g_-\oplus g_+$, $f\in\gotH$, $g_\pm\in\cD_\pm$ and
$g=g_-\oplus g_+\in L^2(\dR,\cH_D)$.
By $J_+$ and $J_-$ we denote the operators
\begin{equation*}
J_+: L^2(\dR,\cH_D)\rightarrow\gotK,\quad g\mapsto 0\oplus 0\oplus g_+,
\end{equation*}
and
\begin{equation*}
J_-: L^2(\dR,\cH_D)\rightarrow\gotK,\quad g\mapsto 0\oplus g_-\oplus 0,
\end{equation*}
respectively.
Observe that $J_++J_-$ is the embedding of $L^2(\dR,\cH_D)$ into $\gotK$.
The subspaces $\cD_+$ and $\cD_-$ are so-called
{\it outgoing} and {\it incoming subspaces} for the selfadjoint dilation
$\widetilde K$ in $\gotK$, that is, one has
\begin{equation*}
e^{-it\widetilde K} \cD_\pm \subseteq\cD_\pm,\,\,\,\,t\in\dR_\pm,  \quad\text{and}\quad
\bigcap_{t\in \bR}e^{-it\widetilde K}\kD_\pm = \{0\}.
\end{equation*}
If, in addition, $\sigma(A_0)$ is singular, then
\begin{equation}\label{3.27x}
\overline{\bigcup_{t\in \bR}e^{-it\widetilde K}\kD_+}= \overline{\bigcup_{t\in
\bR}e^{-it\widetilde K}\kD_-}=\gotK^{ac}(\widetilde{K})
\end{equation}
holds.
Hence $\{\widetilde K,\cD_-,\cD_+\}$ is a Lax-Phillips scattering system
and, in particular,
the {\it Lax-Phillips wave operators}
\begin{equation}\label{3.27xx}
\gO_\pm := \slim_{t\to\pm\infty}e^{it\widetilde K}J_\pm
e^{-itG_0}:L^2(\dR,\cH_D)\rightarrow\gotK
\end{equation}
exist, cf. \cite{BW}.
Since $\slim_{t\to\pm\infty}J_\mp e^{-itG_0} = 0$
the restrictions of the wave operators $W_\pm(\widetilde K,K_0)$
of the scattering system $\{\widetilde K,K_0\}$ onto $L^2(\dR,\cH_D)$
coincide with the Lax-Phillips wave operators $\Omega_\pm$,
\begin{equation*}
W_\pm(\widetilde K,K_0)\iota_{L^2} =
\slim_{t\to\pm\infty}e^{it\widetilde K}(J_++J_-) e^{-itG_0} = \Omega_\pm.
\end{equation*}
%
%
%%coincide with the Lax-Phillips wave operators $\Omega_\pm$; 
Here $\iota_{L^2}$ is the canonical embedding of 
$L^2(\dR,\cH_D)$ into $\sK$.
Hence the {\it Lax-Phillips scattering operator} $S^{LP} := \gO^*_+\gO_-$
admits the representation
\begin{equation*}
S^{LP}= P_{L^2}S(\widetilde K,K_0)\,\iota_{L^2}
\end{equation*}
where $S(\widetilde K,K_0)= W_+(\widetilde K,K_0)^*W_-(\widetilde K,K_0)$ is the scattering operator
of the scattering system $\{\widetilde K,K_0\}$ and $P_{L^2}$ is the orthogonal projection from $\sK$ onto  $L^2(\dR,\cH_D)$. Hence the Lax-Phillips scattering operator $S^{LP}$
is a contraction in $L^2(\dR,\cH_D)$ and commutes with the selfadjoint differential operator $G_0$.
Therefore $S^{LP}$ is unitarily equivalent to a multiplication operator
induced by a family $\{S^{LP}(\lambda)\}_{\lambda\in\dR}$ of contractive operators in $L^2(\dR,\cH_D)$; this family
is called the {\it Lax-Phillips scattering matrix}.

The above considerations together with Theorem~\ref{dilscat} immediately
imply the following corollary on the representation of the Lax-Phillips scattering matrix, cf. \cite[Corollary 3.10]{BeMN2}.
\begin{cor}\label{lax1}
Let $\{\widetilde K,\cD_-,\cD_+\}$ be the Lax-Phillips scattering
system considered above and
let $A$, $\Pi=\{\cH,\Gamma_0,\Gamma_1\}$, $A_D$, $M(\cdot)$ and $G_0$ be as
in the beginning of this section. Then the following holds:
\begin{itemize}
 \item [{\rm (i)}]  $G_0 =G_0^{ac}$ is unitarily
equivalent to the multiplication operator
with the free variable in
$L^2(\dR,\cH_D)=L^2(\dR,d\lambda,\kH_D)$.
\item [{\rm (ii)}] In $L^2(\dR,d\lambda,\kH_D)$ the Lax-Phillips scattering matrix
$\{S^{LP}(\gl)\}_{\lambda\in\dR}$ admits the representation
\begin{equation}\la{5.11}
S^{LP}(\lambda)= I_{\kH_D}+2i\sqrt{-\imag(D)}\bigl(D-M(\lambda)\bigr)^{-1}
\sqrt{-\imag(D)}
\end{equation}
for a.e. $\lambda\in\bR$, where $M(\lambda)=M(\lambda+i0)$.
\end{itemize}
\end{cor}
Let again $A_D$ be the maximal dissipative extension of $A$ corresponding to
the maximal dissipative matrix $D\in[\cH]$ and let $\cH_D=\ran(\imag (D))$. By
\cite{DM92} the characteristic function $W_{A_D}(\cdot)$ of the 
$A_D$ is given by
\begin{equation}\label{2.81}
\begin{split}
W_{A_D} :\dC_-&\rightarrow [\cH_D]\\
\mu\mapsto I_{\kH_D}&
-2i\sqrt{-\imag(D)}\bigl(D^* - M(\mu)\bigr)^{-1}\sqrt{-\imag(D)}.
\end{split}
\end{equation}
It determines a completely non-selfadjoint part of $A_D$ uniquely up to unitary 
equivalence.

Comparing \eqref{5.11} and \eqref{2.81} we obtain the famous relation
between the Lax-Phillips scattering matrix and the characteristic function discovered
originally  by  Adamyan and Arov in \cite{AA1,AA2,AA3,AA4}, cf. \cite[Corollary 3.11]{BeMN2}
for another proof and further development.
\begin{cor}\label{adamyanarov}
Let the assumption be as in Corollary~{\rm \ref{lax1}}. Then the Lax-Phillips scattering matrix
$\{S^{LP}(\lambda)\}_{\lambda\in\dR}$ and the characteristic function $W_{A_D}(\cdot)$ of the
maximal dissipative operator $A_D$ are related by
\begin{equation*}
S^{LP}(\gl) = W_{A_D}(\gl - i0)^*
\end{equation*}
for a.e $\lambda\in\bR$. 
\end{cor}

\subsection{A modified Birman-Krein formula for dissipative scattering systems}\label{modbk}

Let $\{\widetilde K,K_0\}$ be the complete scattering system from the previous subsections and let $\{\widetilde S(\lambda)\}_{\lambda\in\dR}$ be the corresponding
scattering matrix. If $\xi_{\widetilde\Theta}(\cdot)$ is the spectral shift function in \eqref{xiwt}, then the
Birman-Krein formula
\begin{equation}
\det(S_{\widetilde{\gT}}(\gl)) = \exp\bigl(-2\pi i \xi_{\widetilde{\gT}}(\gl)\bigr)
\end{equation}
holds for a.e.  $\gl \in \bR$, see Theorem \ref{V.1a}. In the next theorem we prove a variant of the Birman-Krein formula for dissipative scattering systems.
\bt\la{t.III.10}
Let $A$ and $G$ be the symmetric operators from Section~\ref{sec3.1} and let $\Pi=\{\cH,\Gamma_0,\Gamma_1\}$ be a boundary triplet for $A^*$ with Weyl function $M(\cdot)$.
Let $D\in[\cH]$ be dissipative and let
$A_D=A^*\!\upharpoonright\ker(\Gamma_1-D\Gamma_0)$ be the corresponding maximal dissipative extension of $A$ 
Then the spectral
shift function $\eta_D(\cdot)$ of the pair $\{A_D,A_0\}$ given by
\eqref{t.4.16} and the scattering matrices $\{S_D(\gl)\}_{\lambda\in\dR}$ and
$\{S^{LP}(\gl)\}_{\lambda\in\dR}$ from Corollary~\ref{disscatcor} and Corollary~{\rm\ref{lax1}} are related via
\begin{equation}\la{t.3.38}
\det(S_D(\gl)) = \overline{\det(S^{LP}(\gl))}\exp\bigl(-2\pi i \eta_D(\gl)\bigr)
\end{equation}
and
\begin{equation}\label{polk}
\det(S^{LP}(\gl)) = \overline{\det(S_D(\gl))}\exp\bigl(-2\pi i \eta_D(\gl)\bigr)
\end{equation}
for a.e. $\gl \in \bR$.
\et
\begin{proof}
Let $\widetilde K$ be the minimal selfadjoint dilation of $A_D$ from \eqref{widetildek} corresponding to
the selfadjoint parameter $\widetilde\Theta$ in \eqref{wtdef} via the boundary triplet $\widetilde\Pi=\{\widetilde\cH,\widetilde\Gamma_0,\widetilde\Gamma_1\}$.
Taking into account Corollary \ref{t.II.2} it follows that the scattering matrix $\{\widetilde S(\lambda)\}_{\lambda\in\dR}$ of the scattering system $\{\widetilde K,K_0\}$ 
satisfies
\begin{equation}\label{ssop}
\det(\widetilde S(\gl)) = \det(\widetilde S_{\widetilde{\gT}_{\rm op}}(\gl)),
\end{equation}
where $\widetilde\Theta_\op$ is the operator part of $\widetilde\Theta$ from \eqref{wtdefop} and
\begin{equation*}
\widetilde S_{\widetilde{\gT}_{\rm op}}(\gl) =
I_{\widetilde{\kH}_{\widetilde M_\op\!(\lambda)}} + 2i\sqrt{\imag(\widetilde{M}_{\rm op}(\gl))}
\bigl(\widetilde{\gT}_{\rm op} - \widetilde{M}_{\rm op}(\gl)\bigr)^{-1}
\sqrt{\imag(\widetilde{M}_{\rm op}(\gl))}
\end{equation*}
for a.e. $\lambda\in\dR$. Making use of $\widetilde{\gT}_{\rm op} = 0$ (see \eqref{wtdefop}) and formula \eqref{t.4.21} we
obtain
\begin{equation*}
\begin{split}
\det\bigl(\widetilde S_{\widetilde{\gT}_\op}(\gl)\bigr) & = 
\det\Bigl(I_{\widetilde{\kH}_{\widetilde M_\op\!(\lambda)}} + 
2i\imag\bigl(\widetilde{M}_{\rm op}(\gl)\bigr)\bigl(\widetilde{\gT}_{\rm op} - \widetilde{M}_{\rm op}(\gl)\bigr)^{-1}\Bigr)\\
& = 
\det\bigl (I_\kH -2i\imag(M(\gl) -D)(M(\gl) - D)^{-1}\bigr)\\
& = 
\frac{\det(M(\gl)^* - D^*)}{\det(M(\gl) - D)}.
\end{split}
\end{equation*}
Hence
\begin{equation*}
\frac{\det(M(\gl)^* - D)}{\det(M(\gl)^* - D^*)}\det\bigl(\widetilde S_{\widetilde{\gT}_{\rm op}}(\gl)\bigr) 
= \frac{\det(M(\gl)^* - D)}{\det(M(\gl) - D)}.
\end{equation*}
Obviously we have
\begin{equation*}
\frac{\det(M(\gl)^* - D)}{\det(M(\gl)^* - D^*)}
= \det\bigl(I_\cH -2i\imag(D)(M(\gl)^* - D^*)^{-1}\bigr)
\end{equation*}
and since
\begin{equation*}
\begin{split}
& \det\bigl(I_\cH -2i\imag(D)(M(\gl)^* - D^*)^{-1}\bigr)\\
&\qquad = 
\det\bigl(I_\cH -2i\sqrt{-\imag(D)}(D^* - M(\gl)^*)^{-1}\sqrt{-\imag(D)}\bigr) \\
& \qquad = \overline{\det(S_{LP}(\gl))}
\end{split}
\end{equation*}
we get
\begin{equation*}
\frac{\det(M(\gl)^* - D)}{\det(M(\gl)^* - D^*)}\det\bigl(\widetilde S_{\widetilde{\gT}_{\rm op}}(\gl)\bigr) 
= \overline{\det(S_{LP}(\gl))}\det\bigl(\widetilde S_{\widetilde{\gT}_{\rm op}}(\gl)\bigr).
\end{equation*}
Similarly, we find
\begin{equation*}
\begin{split}
&\frac{\det(M(\gl)^* - D)}{\det(M(\gl) - D)} \\
&\qquad \,\,\,=\det\bigl(I_\cH + 2i\sqrt{\imag(M(\gl)}(D - M(\gl))^{-1}\sqrt{\imag(M(\gl)}\,\bigr) \\
&\qquad \,\,\,= \det(S_D(\gl)),
\end{split}
\end{equation*}
so that the relation
\begin{equation*}
 \overline{\det(S_{LP}(\gl))}\det\bigl(\widetilde S_{\widetilde{\gT}_{\rm op}}(\gl)\bigr)= \det(S_D(\gl))
\end{equation*}
holds for a.e. $\lambda\in\dR$. Hence the Birman-Krein formula 
\begin{equation*}
\det\bigl(\widetilde S(\lambda)\bigr)=\exp\bigl(-2\pi i\xi_{\widetilde\Theta}(\lambda)\bigr),
\end{equation*}
which connects the scattering matrix of $\{\widetilde K,K_0\}$ and the spectral shift function $\xi_{\widetilde\Theta}(\cdot)$ in \eqref{xiwt}, 
Theorem~\ref{III.8} and
\eqref{ssop} immediately imply \eqref{t.3.38} and \eqref{polk} for a.e. $\lambda\in\dR$.
\end{proof}

\section{Coupled scattering systems}\la{IV}

In the following we investigate so-called coupled scattering systems in a similar form as in \cite{BeMN2}, where, roughly speaking, the fixed dissipative
scattering system in the previous section is replaced by a family of dissipative scattering systems which can be regarded as an open quantum system. 
These maximal dissipative operators form a \v{S}traus family of extensions of a symmetric operator and their resolvents coincide pointwise with the resolvent of a certain selfadjoint operator
in a bigger Hilbert space. The spectral shift functions of the dissipative scattering systems are explored and a variant of the Birman-Krein formula is proved.

\subsection{\v{S}traus family and coupling of symmetric operators}\label{IV.1}

Let $A$ be a densely defined closed simple symmetric operator 
with equal finite deficiency indices $n_\pm(A)$ in the separable
Hilbert space $\gotH$ and let $\gP_A = \{\kH,\gG_0,\gG_1\}$ be a boundary triplet for $A^*$ with $\gamma$-field $\gamma(\cdot)$ and Weyl function $M(\cdot)$.
Furthermore, 
let $T$ be a densely defined closed simple symmetric operator with equal finite deficiency indices $n_\pm(T)=n_\pm(A)$ in the
separable Hilbert space $\gotG$ and let $\gP_T = \{\kH,\gY_0,\gY_1\}$ be a boundary triplet of $T^*$ with 
$\gga$-field $\nu(\cdot)$ and Weyl function $\gt(\cdot)$. 

Observe that $-\tau(\lambda)\in[\cH]$ is a dissipative matrix for each $\lambda\in\dC_+$ and therefore by Proposition~\ref{propo}
\begin{equation}\label{strausfam}
A_{-\gt(\gl)} := A^*\upharpoonright\ker\bigl(\gG_1 + \gt(\gl)\gG_0\bigr), \quad
\gl \in \bC_+,
\end{equation}
is a family of maximal dissipative extensions of $A$ in $\sH$. This family is called the {\it \v{S}traus family of $A$ associated with $\tau$}.
Since the limit $\gt(\gl) :=
\gt(\gl + i0)$ exists for a.e. $\gl \in \bR$ the \v{S}traus family admits
an extension to the real axis for a.e. $\gl \in \bR$. Analogously the 
\v{S}traus family 
\begin{equation}\label{strausfam2}
T_{-M(\lambda)} := T^*\upharpoonright\ker\bigl(\gY_1 + M(\gl)\gY_0\bigr), \quad
\gl \in \bC_+,
\end{equation}
of $T$ associated with $M$
consists of maximal dissipative extensions of $T$ in $\sG$ and admits
an extension to the real axis for a.e. $\gl \in \bR$. Sometimes it is convenient to define the \v{S}traus family also
on $\dC_-$, in this case the extensions $A_{-\gt(\gl)}$ and $T_{-M(\lambda)}$ are maximal accumulative 
for $\lambda\in\dC_-$, cf. Proposition~\ref{propo}.

In a similar way as in Section~\ref{sec3.1} we consider the densely defined closed simple symmetric operator
\begin{equation*}
L:=\begin{pmatrix} A & 0 \\ 0 & T\end{pmatrix}
\end{equation*}
with equal finite deficiency indices $n_\pm(L)=2 n_\pm(A)=2 n_\pm(T)$ in the separable Hilbert space
$\sL=\sH\oplus\sG$. Then obviously $\gP_L =
\{\widetilde{\kH},\widetilde{\gG}_0,\widetilde{\gG}_1\}$, where $\widetilde\cH:=\cH\oplus\cH$
\begin{equation}\label{tildetrip}
\widetilde{\gG}_0(f\oplus g) :=
\begin{pmatrix}
\gG_0 f\\
\gY_0 g
\end{pmatrix}
\quad \mbox{and} \quad 
\widetilde{\gG}_1(f\oplus g)  :=
\begin{pmatrix}
\gG_1 f\\
\gY_1g
\end{pmatrix},
\end{equation}
$f\in\dom (A^*)$, $g\in\dom (T^*)$, is a boundary triplet for the adjoint 
\begin{equation*}
L^*=\begin{pmatrix} A^* & 0 \\ 0 & T^*\end{pmatrix}.
\end{equation*} 
The $\gga$-field $\widetilde{\gga}(\cdot)$ and
Weyl function $\widetilde{M}(\cdot)$ corresponding to the boundary triplet 
$\gP_L =
\{\widetilde{\kH},\widetilde{\gG}_0,\widetilde{\gG}_1\}$ are given by 
\begin{equation*} 
\widetilde\gamma(\lambda)=\begin{pmatrix} \gamma(\lambda) & 0 \\ 0 & \nu(\lambda)\end{pmatrix}\quad
\text{and}\quad\widetilde{M}(\gl) = 
\begin{pmatrix}
M(\gl) & 0 \\
0 & \gt(\gl)
\end{pmatrix},\quad \gl \in \bC \backslash \bR,
\end{equation*}
cf. \eqref{3.14a} and \eqref{3.14aa}. In the sequel we investigate the scattering system consisting of
the selfadjoint operator
\begin{equation}\label{l0}
L_0 :=
L^*\upharpoonright \ker(\widetilde{\gG}_0)=\begin{pmatrix} A_0 & 0\\ 0 & G_0\end{pmatrix},
\end{equation}
where $A_0 = A^*\!\upharpoonright\!\ker(\gG_0)$ and $T_0=T^*\!\upharpoonright\!\ker(\gY_0)$, 
and the selfadjoint operator $\widetilde{L} =L^*\upharpoonright \widetilde{\gG}^{-1}\gT$ which corresponds to
the selfadjoint relation 
\begin{equation}\label{wttheta}
\gT:=\left\{
\begin{pmatrix} (v,v)^\top\\ (w,-w)^\top
\end{pmatrix}:
v,w\in\kH\right\}
\end{equation}
in $\widetilde\cH$. The selfadjoint extension $\widetilde L$ of $L$ is sometimes called a coupling
of the
subsystems $\{\gotH,A\}$ and $\{\gotG,T\}$, cf. \cite{DHMS00}. In the following theorem $\widetilde L$
and its connection to the \v{S}traus family in \eqref{strausfam} is made explicit, cf. \cite{BeMN2,DHMS00}.
\begin{thm}\label{coupling}
Let $A$, $\Pi_A=\{\cH,\Gamma_0,\Gamma_1\}$, $M(\cdot)$, $T$, $\gP_T = \{\kH,\gY_0,\gY_1\}$, $\gt(\cdot)$ and 
$L$ be as above.
Then the selfadjoint extension $\widetilde{L}$ of $L$ in $\sL$ is given by
\begin{equation}\label{athe}
\widetilde L=L^*\!\upharpoonright \left\{
f \oplus g \in\dom(L^*):
\begin{array}{l}
\Gamma_0 f-\gY_0 g = 0\\
\Gamma_1 f+\gY_1 g=0
\end{array}
\right\}
\end{equation}
and satisfies
\begin{displaymath}
P_\gotH\bigl(\widetilde L-\lambda)^{-1}\upharpoonright_{\gotH}=
\bigl(A_{-\tau(\lambda)}-\lambda\bigr)^{-1}\,\,\,\text{and}\,\,\,P_\gotG\bigl(\widetilde L-\lambda)^{-1}\upharpoonright_{\gotG}=
\bigl(T_{-M(\lambda)}-\lambda\bigr)^{-1}
\end{displaymath}
for all $\lambda\in \dC\backslash\dR$. Moreover, the following minimality conditions hold:
\begin{equation*}
\gotL =\clospa\bigl\{\bigl(\widetilde L-\lambda\bigr)^{-1}\gotH:
\lambda\in\bC\backslash\bR\bigr\}=\clospa\bigl\{\bigl(\widetilde L-\lambda\bigr)^{-1}\gotK:
\lambda\in\bC\backslash\bR\bigr\}.
\end{equation*}
\end{thm}

\subsection{Spectral shift function and trace formula for a coupled scattering system}

Next we calculate the spectral shift function of the complete scattering system 
$\{\widetilde{L},L_0\}$. By Theorem \ref{V.1} a spectral shift
function $\widetilde{\xi}_\gT(\cdot)$ is given by
\begin{equation}\label{zxc}
\widetilde{\xi}_\Theta(\gl) =
\frac{1}{\pi}\imag\bigl(\tr\bigl(\log(\widetilde{M}_{\rm op}(\gl+i0) - \gT_{\rm op})\bigr)\bigr)
\end{equation}
for a.e. $\gl \in \bR$, where 
\begin{equation}\label{theop}
\gT_{\rm op}:=\left\{
\begin{pmatrix} (v,v)^\top\\ (0,0)^\top
\end{pmatrix}:
v\in\kH\right\}
\end{equation}
is the operator part of $\Theta$ in the space 
\begin{equation}\label{theop2}
\widetilde\cH_\op:=\left\{
\begin{pmatrix} v \\ v \end{pmatrix}: v\in\cH \right\}\subset\widetilde\cH
\end{equation}
and $\widetilde M_\op(\cdot)=\widetilde P_\op\widetilde M(\cdot)\widetilde\iota_\op$ 
denotes compression of the Weyl function $\widetilde M(\cdot)$ in $\widetilde\cH$
onto $\widetilde\cH_\op$. Observe that $\Theta_\op=0$ so that the spectral shift function 
$\widetilde\xi_\Theta(\cdot)$ in \eqref{zxc}
has the form
\begin{equation}\label{t.4.8}
\widetilde{\xi}_\Theta(\gl) =
\frac{1}{\pi}\imag\bigl(\tr\bigl(\log(\widetilde{M}_{\rm op}(\gl+i0))\bigr)\bigr)
\end{equation}
for a.e. $\gl \in \bR$. Furthermore, the trace formula
\begin{equation}\la{t.4.7}
\tr\bigl((\widetilde{L} - \gl)^{-1} - (L_0 - \gl)^{-1}\bigr) =
-\int_\bR \frac{1}{(t-\gl)^{2}}\,\widetilde{\xi}_\Theta(t)\,dt
\end{equation}
holds for all $\gl \in \bC \backslash \bR$. 
\bt\la{t.IV.2}
Let $A$, $\Pi_A=\{\cH,\Gamma_0,\Gamma_1\}$, $M(\cdot)$ and
$T$, $\gP_T = \{\kH,\gY_0,\gY_1\}$, $\gt(\cdot)$ be as in
the beginning of Section \ref{IV.1}.
Then the spectral shift function $\widetilde{\xi}_\Theta(\cdot)$ of the pair
$\{\widetilde{L},L_0\}$ admits the representation
\begin{equation}\la{t.4.9}
\widetilde{\xi}_\Theta(\gl) =
\frac{1}{\pi}\imag\bigl(\tr\bigl(\log(M(\gl+i0) + \gt(\gl+i0))\bigr)\bigr)+2k
\end{equation}
for some $k\in\dZ$ and a.e. $\gl \in \bR$. Moreover, the
modified trace formula
\begin{equation}\la{t.4.10}
\begin{split}
&\tr\left((A_{-\gt(\gl)} - \gl)^{-1} - (A_0 - \gl)^{-1}\right) +\\
&\qquad
\tr\left((T_{-M(\gl)} - \gl)^{-1} - (T_0 - \gl)^{-1}\right) =
-\int_\bR \frac{1}{(t-\gl)^{2}}\,\widetilde{\xi}_\Theta(t)\, dt
\end{split}
\end{equation}
holds for all $\gl \in \bC\backslash\dR$. 
\et
\begin{proof}
With the help of the unitary operator
\begin{equation}\label{tzuv}
\widetilde V:\cH\longrightarrow\widetilde\cH_\op,\qquad x\mapsto\frac{1}{\sqrt{2}}
\begin{pmatrix} x\\ x\end{pmatrix},
\end{equation}
we obtain
\begin{equation}\label{tzu}
\widetilde V^*\widetilde{M}_{\rm op}(\gl)\widetilde V = \frac{1}{2}\left(M(\gl) + \gt(\gl)\right).
\end{equation}
We conclude in the same way as in the proof of Theorem~\ref{III.8} that the functions 
$\tr(\log(\widetilde M_\op(\cdot)))$ and $\tr(\log(M(\cdot)+\tau(\cdot)))$ differ by a constant
and
\begin{equation*}
\exp\bigl(\tr\bigl(\log(\widetilde M_\op(\lambda))\bigr)\bigr)=
\exp\bigl(\tr\bigl(\log(M(\lambda)+\tau(\lambda))\bigr)\bigr)\det \tfrac{1}{2}I_\cH
\end{equation*}
implies that there exists $k\in\dZ$ such that
\begin{equation*}
\imag\bigl(\tr\bigl(\log(\widetilde M_\op(\lambda))\bigr)\bigr)=
\imag\bigl(\tr\bigl(\log(M(\lambda)+\tau(\lambda))\bigr)\bigr)+2k\pi
\end{equation*}
holds. This together with \eqref{t.4.8} implies \eqref{t.4.9}.

In order to verify the trace formula \eqref{t.4.10} note that by \eqref{resoll} we have
\begin{equation}
(\widetilde{L} - \gl)^{-1} - (L_0 - \gl)^{-1} = 
\widetilde{\gga}(\gl)
\bigl(\widetilde{\gT} - \widetilde{M}(\gl)\bigr)^{-1}
\widetilde{\gga}(\bar{\gl})^*
\end{equation}
for all $\lambda\in\rho(\widetilde L)\cap\rho(L_0)$.
Taking into account \eqref{t.2.5} we get
\begin{equation}
(\widetilde{L} - \gl)^{-1} - (L_0 - \gl)^{-1} = 
-\widetilde{\gga}(\gl)\widetilde\iota_\op
\bigl(\widetilde M_{\rm op}(\gl)\bigr)^{-1}\widetilde P_{\rm op}
\widetilde{\gga}(\bar{\gl})^*
\end{equation}
and by using 
\begin{equation*}
\bigl(\widetilde M_\op(\lambda)\bigr)^{-1}=
2\widetilde V\bigl((M(\lambda)+\tau(\lambda)\bigr)^{-1}\widetilde V^*,
\end{equation*}
cf. \eqref{tzu},
we obtain
\begin{equation}
(\widetilde{L} - \gl)^{-1} - (L_0 - \gl)^{-1} = 
-2\widetilde{\gga}(\gl)\widetilde\iota_\op
\widetilde V\bigl(M(\gl) + \gt(\gl)\bigr)^{-1}\widetilde V^*\widetilde P_\op
\widetilde{\gga}(\bar{\gl})^*
\end{equation}
which yields
\begin{equation}
\tr\bigl((\widetilde{L} - \gl)^{-1} - (L_0 - \gl)^{-1}\bigr) = 
-2\tr\bigl(
\bigl(M(\gl) + \gt(\gl)\bigr)^{-1}\widetilde V^*\widetilde P_\op
\widetilde{\gga}(\bar{\gl})^*\widetilde{\gga}(\gl)\widetilde\iota_\op\widetilde V\bigr)
\end{equation}
for all $\lambda\in\rho(\widetilde L)\cap\rho(L_0)$. As in \eqref{mlambda} we find  
\begin{equation}
\widetilde P_\op\widetilde{\gga}(\bar{\gl})^*\widetilde{\gga}(\gl)\widetilde\iota_\op
=\widetilde P_\op\frac{d}{d\gl}\widetilde{M}(\gl)\widetilde\iota_\op=\frac{d}{d\gl}\widetilde{M}_\op(\gl)
\end{equation}
and with the help of \eqref{tzu} we conclude
\begin{equation}
\widetilde V^*\widetilde P_\op \widetilde{\gga}(\bar{\gl})^*\widetilde{\gga}(\gl)\widetilde\iota_\op\widetilde V =
\frac{1}{2}\left(\frac{d}{d\gl}M(\gl) + \frac{d}{d\gl}\gt(\gl)\right).
\end{equation}
Hence
\begin{equation*}
\begin{split}
&\tr\bigl((\widetilde{L} - \gl)^{-1} - (L_0 - \gl)^{-1}\bigr) \\
&\quad \quad=
-\tr\left(\bigl(M(\gl) + \gt(\gl)\bigr)^{-1}\left(\frac{d}{d\gl}M(\gl) +
  \frac{d}{d\gl}\gt(\gl)\right)\right).
\end{split}
\end{equation*}
Using again \eqref{mlambda} we find
\begin{equation*}
\begin{split}
&\tr\bigl((\widetilde{L} - \gl)^{-1} - (L_0 - \gl)^{-1}\bigr) \\
& \quad =
-\tr\left(\gga(\gl)(M(\gl) + \gt(\gl))^{-1}\gga(\bar{\gl})^*\right) 
-\tr\left(\nu(\gl)(M(\gl) + \gt(\gl))^{-1}\nu(\bar{\gl})^*\right).
\end{split}
\end{equation*}
By \eqref{resoll} the resolvents of the \v{S}traus family of $A$ associated with $\tau$
and the \v{S}traus family of $T$ associated with $M$ are given by
\begin{equation}\la{t.4.24}
\bigl(A_{-\gt(\gl)} - \gl\bigr)^{-1} - (A_0 - \gl)^{-1} = 
-\gga(\gl)\bigl(M(\gl) + \gt(\gl)\bigr)^{-1}\gga(\bar{\gl})^*
\end{equation}
and
\begin{equation}\la{t.4.25}
\bigl(T_{-M(\gl)} - \gl\bigr)^{-1} - (T_0 - \gl)^{-1} =
-\nu(\gl)\bigl(M(\gl) + \gt(\gl)\bigr)^{-1}\nu(\bar{\gl})^*,
\end{equation}
respectively.
Taking into account \eqref{t.4.24}, \eqref{t.4.25} and \eqref{t.4.7}
we prove \eqref{t.4.10}. 
\end{proof}
Let us consider the the spectral shift function 
$\eta_{-\gt(\mu)}(\cdot)$ of the dissipative scattering system
$\{A_{-\gt(\mu)},A_0\}$ for those $\mu
\in \bR$ for which the limit $\gt(\mu) := \gt(\mu + i0)$ exists.
By Theorem \ref{III.8} the function $\eta_{-\gt(\mu)}(\cdot)$ admits the
representation
\begin{equation}\label{etatau}
\eta_{-\gt(\mu)}(\lambda) = \frac{1}{\pi}\imag\bigl(\tr\bigl(\log(M(\lambda+i0) +
    \gt(\mu))\bigr)\bigr)
\end{equation}
for a.e. $\lambda\in\dR$. Moreover, we have
\begin{equation*}
\tr\left((A_{-\gt(\mu)} - \lambda)^{-1} - (A_0 - \lambda)^{-1}\right) =
-\int_\bR \frac{1}{(t - \lambda)^2}\,\eta_{-\gt(\mu)}(t)\;dt
\end{equation*}
for all $\lambda \in \bC_+$, cf. Theorem~\ref{III.8}. Similarly, we introduce
the spectral shift function $\eta_{-M(\mu)}(\cdot)$ of the dissipative scattering system
$\{T_{-M(\mu)},T_0\}$ for those $\mu\in\dR$ for which the limit $M(\mu)=M(\mu+i0)$ exists. It follows that 
\begin{equation}\label{etam}
\eta_{-M(\mu)}(\lambda) = \frac{1}{\pi}\imag\bigl(\tr\bigl(\log(M(\mu) +
    \gt(\lambda+i0))\bigr)\bigr)
\end{equation}
holds for a.e. $\gl \in \dR$ and 
\begin{equation*}
\tr\left((T_{-M(\mu)} - \lambda)^{-1} - (T_0 - \lambda)^{-1}\right) =
-\int_\bR \frac{1}{(t - \lambda)^2}\,\eta_{-M(\mu)}(t)\;dt
\end{equation*}
is valid for $\lambda \in \bC_+$. Hence we get
immediately the following corollary.
\bc\la{t.IV.3}
Let the assumptions be as in Theorem~\ref{t.IV.2}, let $L_0$, $\widetilde L$ be as in \eqref{l0}, \eqref{athe} 
and let $\eta_{-\gt(\mu)}(\cdot)$ and 
$\eta_{-M(\mu)}(\cdot)$ be the spectral shift functions in \eqref{etatau} and \eqref{etam}, respectively.
Then the spectral shift function $\widetilde{\xi}_\Theta(\cdot)$ of the pair
$\{\widetilde{L},L_0\}$ admits the representation
\begin{equation}
\widetilde{\xi}_\Theta(\gl) = \eta_{-\tau(\gl)}(\gl)+2k = \eta_{-M(\gl)}(\gl)+2l
\end{equation}
for a.e. $\lambda\in\dR$ and some $k,l\in\dZ$.
\ec

\subsection{Scattering matrices of coupled systems}\label{4.3sec}

We investigate the scattering matrix of the scattering system $\{\widetilde{L},L_0\}$, where
$\widetilde L$ and $L_0$ are the selfadjoint operators in $\sL=\sH\oplus\sG$ 
from \eqref{athe} and \eqref{l0}, respectively.   
By Theorem \ref{scattering} the scattering matrix
$\{\widetilde{S}_\Theta(\gl)\}_{\gl \in \bR}$ of 
$\{\widetilde{L},L_0\}$ admits the representation
\begin{equation}\la{t.4.26}
\widetilde S_\gT(\gl) = I_{\widetilde{\kH}_{\widetilde{M}(\gl)}} + 2 i 
\sqrt{\imag(\widetilde{M}(\gl))}\bigl(\gT -
\widetilde{M}(\gl)\bigr)^{-1}\sqrt{\imag(\widetilde{M}(\gl))}. 
\end{equation}
Here $\widetilde M(\cdot)$ is the Weyl function of the boundary triplet $\Pi_L=
\{\widetilde\cH,\widetilde\Gamma_0,\widetilde\Gamma_1\}$ from \eqref{tildetrip} and 
\begin{equation*}
\widetilde{\kH}_{\widetilde{M}(\gl)}:=\ran \bigl(\imag \bigl( \widetilde M(\lambda+i0)\bigr)\bigr)
\end{equation*} 
for a.e. $\lambda\in\dR$.
In \cite{BeMN2} the scattering matrix of $\{\widetilde{L},L_0\}$ 
was expressed in terms of the Weyl functions $M(\cdot)$ and $\gt(\cdot)$ of the boundary
triplets $\Pi_A=\{\cH,\Gamma_0,\Gamma_1\}$ and $\Pi_T=\{\cH,\gY_0,\gY_1\}$, respectively.
The following representation for $\{\widetilde S_\gT(\lambda)\}_{\lambda\in\dR}$
can be deduced from Corollary \ref{t.II.2}.

\begin{thm}\label{scatcoup}
Let $A$, $\Pi_A=\{\cH,\Gamma_0,\Gamma_1\}$, $M(\cdot)$ and
$T$, $\gP_T = \{\kH,\gY_0,\gY_1\}$, $\gt(\cdot)$ be as above.
Then the following holds:
\begin{itemize}
\item [{\rm (i)}] $L_0^{ac}=A^{ac}_0\oplus T_0^{ac}$ 
is unitarily equivalent to the multiplication operator
with the free variable in $L^2(\bR,d\gl,\kH_{M(\gl)}\oplus\cH_{\tau(\lambda)})$.

\item [\;\;{\rm (ii)}] In $L^2(\bR,d\gl,\kH_{M(\gl)}\oplus\cH_{\tau(\lambda)})$ the scattering
matrix $\{\widetilde S_\Theta(\gl)\}_{\lambda\in\dR}$ of the complete scattering system $\{\widetilde L,L_0\}$ is given by
\begin{equation} \label{4.13A}
\widetilde S_\Theta(\gl) = I_{\kH_{M(\gl)}\oplus \kH_{\tau(\lambda)}}
-2i\begin{pmatrix} 
\widetilde T_{11}(\lambda) &  \widetilde T_{12}(\lambda)\\
\widetilde T_{21}(\lambda) &  \widetilde T_{22}(\lambda)
\end{pmatrix}
\in[\cH_{M(\lambda)}\oplus\cH_{\tau(\lambda)}],
\end{equation}
for a.e. $\gl \in \bR$ where
\begin{equation*}
\begin{split}
\widetilde T_{11}(\lambda)&=\sqrt{\imag(M(\lambda))}
\bigl(M(\gl)+\tau(\lambda)\bigr)^{-1}\sqrt{\imag(M(\lambda))},\\
\widetilde T_{12}(\lambda)&=\sqrt{\imag(M(\lambda))}
\bigl(M(\gl)+\tau(\lambda)\bigr)^{-1}\sqrt{\imag(\tau(\lambda))},\\
\widetilde T_{21}(\lambda)&=
\sqrt{\imag(\tau(\lambda))}\bigl(M(\gl)+\tau(\lambda)\bigr)^{-1}
\sqrt{\imag(M(\lambda))},\\
\widetilde T_{22}(\lambda)&=
\sqrt{\imag(\tau(\lambda))}\bigl(M(\gl)+\tau(\lambda)\bigr)^{-1}
\sqrt{\imag(\tau(\lambda))}
\end{split}
\end{equation*}
and $M(\lambda)=M(\lambda+i0)$, $\tau(\lambda)=\tau(\lambda+i0)$.
\end{itemize}
\end{thm}
Let $J_\gotH: \gotH \longrightarrow \sL$ and $J_\gotG:
\gotG \longrightarrow \sL$ 
the natural embedding
operators of the subspaces $\gotH$ and $\gotG$ into
$\sL$, respectively. 
The wave operators
\begin{equation}
W_\pm(\widetilde{L},A_0) :=
\slim_{t\to\pm\infty}e^{it\widetilde{L}}J_\gotH e^{-itA_0}P^{ac}(A_0)
\end{equation}
and
\begin{equation}
W_\pm(\widetilde{L},T_0) :=
\slim_{t\to\pm\infty}e^{it\widetilde{L}}J_\gotG e^{-itT_0}P^{ac}(T_0)
\end{equation}
are called the {\it channel wave operators} or {\it partial wave operators}. The {\it channel
scattering operators} $S_\gotH$ and $S_\gotG$  are defined by
\begin{equation}
S_\gotH := W_+(\widetilde{L},A_0)^*W_-(\widetilde{L},A_0)
\quad \mbox{and} \quad
S_\gotG :=  W_+(\widetilde{L},T_0)^*W_-(\widetilde{L},T_0).
\end{equation}
The channel scattering operators $S_\gotH$ and $S_\gotG$ are contractions in $\sH^{ac}(A_0)$ and $\sG^{ac}(T_0)$ and commute with
$A_0$ and $T_0$, respectively. Hence, there are measurable families of
contractions 
\begin{equation}\label{channel}
\{S_\gotH(\gl)\}_{\gl \in \bR}\qquad\text{and}\qquad \{S_\gotG(\gl)\}_{\gl \in \bR}
\end{equation}
such that the multiplication operators induced by these families in
the spectral representations $L^2(\bR,d\gl,\kH_{M(\gl)})$ and
$L^2(\bR,d\gl,\kH_{\gt(\gl)})$ 
of $A^{ac}_0$ and $T^{ac}_0$, respectively, are unitarily equivalent
to the channel scattering operators $S_\gotH$ and $S_\gotG$. The multiplication operators in \eqref{channel} are called {\it channel scattering matrices}.
\bc\la{t.IV.4}
Let $A$, $\Pi_A=\{\cH,\Gamma_0,\Gamma_1\}$, $M(\cdot)$ and
$T$, $\gP_T = \{\kH,\gY_0,\gY_1\}$, $\gt(\gl)$ be as above.
Then the following holds:

\begin{itemize}
\item [{\rm (i)}] 
$A^{ac}_0$ and $T_0^{ac}$ 
are unitarily equivalent to the multiplication operators
with the free variable in $L^2(\bR,d\gl,\kH_{M(\gl)}$ and
$L^2(\bR,d\gl,\cH_{\tau(\lambda)})$, respectively.

\item[{\rm (ii)}]
In $L^2(\bR,d\gl,\kH_{M(\gl)})$ and $L^2(\bR,d\gl,\cH_{\tau(\lambda)})$ 
the channel scattering matrices $\{S_\gotH(\gl)\}_{\gl \in \bR}$ and
$\{S_\gotG(\gl)\}_{\gl \in \bR}$ are given by
\begin{displaymath}
S_\gotH(\gl) = I_{\kH_{M(\gl)}} - 2i \sqrt{\imag(M(\lambda))}
\bigl(M(\gl)+\tau(\lambda)\bigr)^{-1}\sqrt{\imag(M(\lambda))}
\end{displaymath}
and
\begin{displaymath}
S_{\gotG}(\gl) =
 I_{\kH_{\gt(\gl)}} -2i
\sqrt{\imag(\tau(\lambda))}\bigl(M(\gl)+\tau(\lambda)\bigr)^{-1}
\sqrt{\imag(\tau(\lambda))}
\end{displaymath}
for a.e. $\gl \in \bR$.
\end{itemize}
\ec

\subsection{A modified Birman-Krein formula for coupled scattering systems}

In a similar way as in Section~\ref{modbk} we prove a variant of the Birman-Krein formula for the coupled scattering system $\{\widetilde L,L_0\}$, where $\widetilde L$
and $L_0$ are as in \eqref{athe} and \eqref{l0}, respectively. 
First of all it is clear that the scattering matrix $\{\widetilde S_\Theta(\lambda)\}_{\lambda\in\dR}$ of $\{\widetilde L,L_0\}$ and the spectral shift function 
$\widetilde\xi_\Theta(\cdot)$ from \eqref{t.4.8} are connected via the usual Birman-Krein formula
\begin{equation}\la{t.4.33}
\det\bigl(\widetilde{S}_\Theta(\gl)\bigr) = \exp\bigl(-2\pi i \widetilde{\xi}_\Theta(\gl)\bigr)
\end{equation}
for a.e. $\gl \in \bR$, cf. Theorem \ref{V.1a}. With the help of the channel scattering matrices from \eqref{channel} and Corollary~\ref{t.IV.4} we find the following
modified Birman-Krein formula.
\bt\la{t.IV.5}
Let $A$ and $T$ be as in Section~\ref{IV.1} and let $\{\widetilde L,L_0\}$ be the complete scattering system from above.
Then the spectral shift function $\widetilde\xi_\Theta(\cdot)$ of the pair
$\{\widetilde L,L_0\}$ in \eqref{t.4.8} is related with the 
channel scattering matrices
$\{S_\gotH(\gl)\}_{\gl \in \bR}$ and $\{S_\gotG(\gl)\}_{\gl \in \bR}$ in \eqref{channel} via 
\begin{equation}\la{t.4.34}
\det(S_\gotH(\gl)) = \overline{\det(S_\gotG(\gl))}\exp\bigl(-2\pi i
  \widetilde{\xi}_\Theta(\gl)\bigr)
\end{equation}
and
\begin{equation}\la{t.4.35}
\det(S_\gotG(\gl)) = \overline{\det(S_\gotH(\gl))}\exp\bigl(-2\pi i
  \widetilde{\xi}_\Theta(\gl)\bigr)
\end{equation}
for a.e. $\gl \in \bR$.
\et
\begin{proof}
Let $\{\widetilde S_\Theta(\lambda)\}_{\lambda\in\dR}$ be the scattering matrix of $\{\widetilde L,L_0\}$ from  \eqref{t.4.26}. Making use of  Corollary \ref{t.II.2} we obtain
\begin{equation}\la{t.4.36}
\det\bigl(\widetilde{S}_\Theta(\gl)\bigr) = \det\bigl(\widetilde S_{\gT_{\rm op}}(\gl)\bigr)
\end{equation}
where $\Theta_\op=0\in[\widetilde\cH_\op]$ is the operator part of $\Theta$ in $\widetilde\cH_\op$, cf. \eqref{theop}, \eqref{theop2}, and
$\widetilde S_{\gT_{\rm op}}(\gl)$ is given by
\begin{equation}
\widetilde S_{\gT_{\rm op}}(\gl) = I_{\kH_{\widetilde{M}_{\rm op}(\gl)}}
-2i\sqrt{\imag(\widetilde{M}_{\rm op}(\gl))}\bigl(\widetilde{M}_{\rm op}(\gl)\bigr)^{-1}
\sqrt{\imag(\widetilde{M}_{\rm op}(\gl))}
\end{equation}
for a.e. $\gl \in \bR$. Here $\widetilde M_\op(\cdot)=\widetilde P_\op\widetilde M(\cdot)\widetilde\iota_\op$ is the compression of the Weyl function 
corresponding to the boundary triplet $\Pi_L=\{\widetilde\cH,\widetilde\Gamma_0,\widetilde\Gamma_1\}$ onto the space $\widetilde\cH_\op$. Let $\widetilde V$ be as in \eqref{tzuv}. Then
we have
\begin{displaymath}
\widetilde M_\op(\lambda)=\frac{1}{2}\widetilde V(M(\lambda)+\tau(\lambda))\widetilde V^*\quad\text{and}\quad 
\widetilde M_\op(\lambda)^{-1}=2 \widetilde V(M(\lambda)+\tau(\lambda))^{-1}\widetilde V^*,
\end{displaymath}
cf. \eqref{tzu}, and therefore we get
\begin{equation*}
\det\bigl(\widetilde S_{\gT_{\rm op}}(\gl)\bigr)=\det\bigl(I_\kH - 2i
\imag\bigl(M(\gl) + \gt(\gl)\bigr)\bigl (M(\gl) + \gt(\gl)\bigr)^{-1}\bigr).
\end{equation*}
This yields
\begin{equation}
\det\bigl(\widetilde S_{\gT_{\rm op}}(\gl)\bigr) = 
\frac{\overline{\det(M(\gl) + \gt(\gl))}}{\det(M(\gl) + \gt(\gl))}
\end{equation}
for a.e. $\gl \in \bR$ and hence
\begin{equation}
\frac{\overline{\det(M(\gl) + \gt(\gl)^*)}}
{\overline{\det(M(\gl) + \gt(\gl))}}\det\bigl(\widetilde S_{\gT_{\rm op}}(\gl)\bigr)
= \frac{\det(M(\gl)^* + \gt(\gl))}{\det(M(\gl) + \gt(\gl))}
\end{equation}
for a.e. $\gl \in \bR$. On the other hand, as a consequence of Corollary \ref{t.IV.4} we obtain
\begin{equation}
\det(S_\gotH(\gl)) = \frac{\det(M(\gl)^* + \gt(\gl))}{\det(M(\gl) + \gt(\gl))}
\end{equation}
and
\begin{equation}
\det(S_\gotG(\gl)) = \frac{\det(M(\gl) + \gt(\gl)^*)}{\det(M(\gl) + \gt(\gl))}
\end{equation}
for a.e. $\gl \in \bR$ and therefore we find
\begin{equation}
\overline{\det(S_\gotG(\gl))}\det\bigl(\widetilde S_{\gT_{\rm op}}(\gl)\bigr) =
\det(S_\gotH(\gl))
\end{equation}
for a.e. $\gl \in \bR$. Taking into account \eqref{t.4.33} and
\eqref{t.4.36} we obtain \eqref{t.4.34}. The relation \eqref{t.4.35}
follows from \eqref{t.4.34}.
\end{proof}
Making use of Corollary~\ref{t.IV.3} we obtain the following form for the relations 
\eqref{t.4.34} and \eqref{t.4.35}.
\bc
Let the assumptions be as in Theorem~\ref{t.IV.5} and let $\eta_{-\tau(\mu)}(\cdot)$ and $\eta_{-M(\mu)}(\cdot)$
be as in \eqref{etatau} and \eqref{etam}, respectively. Then the 
channel scattering matrices
$\{S_\gotH(\gl)\}_{\gl \in \bR}$ and $\{S_\gotG(\gl)\}_{\gl \in \bR}$
are connected with the functions $\lambda\mapsto\eta_{-\tau(\lambda)}(\lambda)$ and
$\lambda\mapsto\eta_{-M(\lambda)}(\lambda)$ via
\begin{equation}
\det(S_\gotH(\gl)) = \overline{\det(S_\gotG(\gl))}\exp\bigl(-2\pi i\eta_{-\gt(\gl)}(\gl)\bigr)
\end{equation}
and
\begin{equation}
\det(S_\gotG(\gl)) = \overline{\det(S_\gotH(\gl))}\exp\bigl(-2\pi i\eta_{-M(\gl)}(\gl)\bigr)
\end{equation}
for a.e. $\lambda\in\dR$.
\ec


\begin{thebibliography}{99}


\bibitem{AA1}
Adamjan, V.M., Arov, D.Z.:
{\em On a class of scattering operators and
characteristic operator-functions of contractions},
Dokl. Akad. Nauk SSSR  160 (1965), 9--12.

\bibitem{AA2}
Adamjan, V.M., Arov, D.Z.:
{\em On scattering operators and contraction semigroups in
Hilbert space},
Dokl. Akad. Nauk SSSR 165 (1965), 9--12.

\bibitem{AA3}
Adamjan, V.M., Arov, D.Z.:
{\em Unitary couplings of semi-unitary operators},
Mat. Issled. 1 (1966) vyp. 2, 3--64.

\bibitem{AA4}
Adamjan, V.M., Arov, D.Z.:
{\em Unitary couplings of semi-unitary operators},
Akad. Nauk Armjan. SSR Dokl. 43 (1966) No. 5, 257--263.

\bibitem{AN90}
Adamjan, V.M., Neidhardt, H.:
{\em  On the summability of the spectral shift function 
for pair of contractions and dissipative operators},
 J. Operator Theory  24  (1990),  no. 1, 187--205. 

\bibitem{AdPav}
Adamjan, V.M., Pavlov, B.S.:
{\em Trace formula for dissipative operators},
Vestnik Leningrad. Univ. Mat. Mekh. Astronom.  1979,  no. 2, 5--9, 118. 

\bibitem{AP1}
Adamyan, V.M., Pavlov, B.S.: {\em Null-range potentials and M.G.
Krein's formula for generalized resolvents}, Zap. Nauchn. Semin.
Leningr. Otd. Mat. Inst. Steklova 149 (1986) 7--23 (russian);
translation in J. Sov. Math.  42 no.2 (1988) 1537--1550. 

\bibitem{BW}
Baumg\"artel, H., Wollenberg, M.:
{\em Mathematical Scattering Theory},
Akademie-Verlag, Berlin, 1983.

\bibitem{BeMN1}
Behrndt, J., Malamud, M.M., Neidhardt, H.:
{\em Scattering matrices and Weyl function},
to appear in Proc. London. Math. Soc.

\bibitem{BeMN2}
Behrndt, J., Malamud, M.M., Neidhardt, H.:
{\em Scattering theory for open quantum systems with finite rank coupling},
to appear in  Math. Phys. Anal. Geom.


\bibitem{BK1}
Birman, M.S.; Krein, M.G.: 
{\em On the theory of wave operators and
scattering operators}, Dokl. Akad. Nauk SSSR 144 (1962), 475--478.

\bibitem{BY92a}
Birman, M.Sh.; Yafaev, D.R.:
{\em  Spectral properties of the scattering matrix}, 
Algebra i Analiz  4  (1992),  no. 6, 1--27;  
translation in  St. Petersburg Math. J. 4 (1993),  no. 6, 1055--1079. 

\bibitem{BY92b}
Birman, M.Sh.; Yafaev, D.R.:
{\em  The spectral shift function. The papers of M.G. Kre\u\i n and their further development},
Algebra i Analiz  4  (1992),  no. 5, 1--44;  
translation in  St. Petersburg Math. J. 4  (1993),  no. 5, 833--870 

\bibitem{BMN1}
Brasche, J.F.; Malamud, M.M.; Neidhardt, H.: {\em Weyl function and
spectral properties of selfadjoint extensions}, Integral Equations
Operator Theory 43 (2002), no. 3, 264--289.

\bibitem{D3}
Davies, E.B.:
{\em  Two-channel Hamiltonians and the optical model of nuclear scattering},
Ann. Inst. H. Poincaré Sect. A (N.S.)  29  (1978), no. 4, 395--413.

\bibitem{D2}
Davies, E.B.:
{\em Nonunitary scattering and capture. I. Hilbert space theory},
Comm. Math. Phys.  71  (1980), no. 3, 277--288.

\bibitem{DHMS00}
Derkach, V.A., Hassi, S., Malamud, M.M., de Snoo, H.:
{\em Generalized resolvents of symmetric operators and admissibility},
Methods Funct. Anal. Topology 6 (2000), 24--53.

\bibitem{DHMS06}
Derkach, V.A., Hassi, S., Malamud, M.M., de Snoo, H.:
{\em Boundary relations and their Weyl families},
Trans. Amer. Math. Soc., 358 (2006), 5351--5400.

\bibitem{DM87}
Derkach, V.A.; Malamud, M.M.:
{\em On the Weyl function and Hermite operators with gaps},
Dokl. Akad. Nauk SSSR  293  (1987),  no. 5, 1041--1046.

\bibitem{DM91}
Derkach, V.A.;  Malamud, M.M.: {\em Generalized resolvents and the
boundary value problems for hermitian operators with gaps}, J.
Funct. Anal. 95 (1991), 1--95.

\bibitem{DM92}
Derkach, V.A.,  Malamud, M.M.:
{\em Characteristic functions of linear operators},
Russian Acad. Sci. Dokl. Math. 45 (1992), 417--424.

\bibitem{DM95}
Derkach, V.A.; Malamud, M.M.: {\em The extension theory of hermitian
operators and the moment problem}, J. Math. Sci. (New York) 73
(1995), 141--242.

\bibitem{DS87}
Dijksma, A.; de Snoo, H.:
{\em Symmetric and selfadjoint relations in Krein spaces I},
Operator Theory: Advances and Applications 24, Birkh\"auser, Basel
(1987), 145--166.

\bibitem{Don1}
Donoghue, W.F.: {\em Monotone Matrix Functions and Analytic
Continuation}, Springer, Berlin-New York, 1974.

\bibitem{Gar1}
Garnett, J.B.: {\em  Bounded Analytic Functions}, Academic Press,
New York-London, 1981.

\bibitem{GMN1}
Gesztesy, F.; Makarov, K.A.; Naboko, S.N.:
{\em The spectral shift operator},
in {\em Mathematica results in quantum mechanics}, J. Dittrich,
P. Exner, M. Tater (eds.), Operator Theory: Advances and Application 108, 
Birkh\"{a}user, Basel, 1999, 59--90. 

\bibitem{GM1}
Gesztesy, F.; Makarov, K.A.:
{\em The $\Xi$ operator and its relation to Krein's spectral shift function},
J. Anal. Math.  81  (2000), 139--183.

\bibitem{GM2}
Gesztesy, F.; Makarov, K.A.:
{\em Some applications of the spectral shift operator},  
in {\em Operator theory and its applications}, A.G. Ramm,
P.N. Shivakumar and A.V. Strauss (eds.), Fields Institute Communication
Series 25, Amer. Math. Soc., Providence, RI, 2000, 267--292. 

\bibitem{GK}
Gohberg, I.; Krein, M.G.:
{\em Introduction to the theory of linear nonselfadjoint operators}, 
Translations of Mathematical Monographs, Vol. 18 American Mathematical Society, Providence, R.I. 1969.

\bibitem{GG}
Gorbachuk, V.I.; Gorbachuk, M.L.: {\em  Boundary Value Problems for
Operator Differential Equations}, Mathematics and its Applications
(Soviet Series) 48, Kluwer Academic Publishers Group, Dordrecht,
1991.

\bibitem{Ka1}
Kato, T.: {\em Perturbation Theory for Linear Operators}, Die
Grundlehren der mathematischen Wissenschaften, Band 132,
2nd edition
Springer, Berlin-New York, 1976.

\bibitem{K49}
Krein, M.G.: {\em Basic propositions of the theory of
representations of hermitian operators with deficiency index
$(m,m)$}, Ukrain. Mat. Z. 1 (1949), 3--66.

\bibitem{K62}
Krein, M.G.: {\em On perturbation determinants and a trace formula
for unitary and selfadjoint operators}, Dokl. Akad. Nauk SSSR  144
(1962), 268--271.

\bibitem{LSY}
Langer, H., de Snoo, H., Yavrian, V.A.:
{\em  A relation for the spectral shift function of two self-adjoint extensions}, 
Operator Theory: Advances and Applications 127 (Birkh\"auser, Basel, 2001), 437--445.

\bibitem{LP}
Lax, P.D., Phillips, R.S.:
{\em Scattering Theory},
Academic Press, New York-London 1967.

\bibitem{Ma1}
Martin, Ph.A.:
{\em  Scattering theory with dissipative interactions and time delay}
Nuovo Cimento B (11)  30  (1975), no. 2, 217--238.

\bibitem{Na1}
Naboko, S.N.:
{\em  Wave operators for nonselfadjoint operators and a functional model},
Zap. Nau\v cn. Sem. Leningrad. Otdel. Mat. Inst. Steklov. (LOMI)  69
(1977), 129--135.

\bibitem{Na2}
Naboko, S.N.:
{\em Functional model of perturbation theory and its applications to scattering theory},
Trudy Mat. Inst. Steklov. 147 (1980), 86--114, 203.

\bibitem{N81}
Neidhardt, H.:
{\em Scattering theory of contraction semigroups}.
Report MATH 1981, 5. Akademie der Wissenschaften der DDR, Institut f\"ur Mathematik, Berlin, 1981.

\bibitem{N84}
Neidhardt, H.:
{\em A dissipative scattering theory}.
Oper. Theory Adv. Appl. 14 (1984),
Birkh\"auser Verlag Basel, 1984, 197--212.

\bibitem{N85}
Neidhardt, H.:
{\em A nuclear dissipative scattering theory},
J. Operator Theory 14 (1985), 57-66.

\bibitem{N86}
Neidhardt, H.:
{\em Eine mathematische Streutheorie f\"ur maximal dissipative Operatoren}.
Report MATH, 86-3. Akademie der Wissenschaften der DDR, Institut f\"ur Mathematik, Berlin, 1986.

\bibitem{N87}
Neidhardt, H.: 
{\em Scattering matrix and spectral shift of the nuclear dissipative scattering theory},
Oper. Theory Adv. Appl., 24, Birkhäuser, Basel, 1987, 237-250. 

\bibitem{N88}
Neidhardt, H.: 
{\em Scattering matrix and spectral shift of the nuclear dissipative scattering theory. II},
J. Operator Theory  19  (1988),  no. 1, 43--62. 

\bibitem{P1} Pavlov, B.S.:
{\em Dilation theory and spectral analysis of nonselfadjoint differential operators},
Mathematical programming and related questions
(Proc. Seventh Winter School, Drogobych, 1974),
Theory of operators in linear spaces (Russian),
pp. 3--69, Central. Èkonom. Mat. Inst. Akad. Nauk SSSR, Moscow, 1976.

\bibitem{P2} Pavlov, B.S.: 
{\em Spectral analysis of a dissipative singular Schr\"odinger operator 
in terms of a functional model},
Partial differential equations, VIII,  87--153, Encyclopaedia
Math. Sci., 65, Springer, Berlin, 1996.

\bibitem{Pe1}
Peller, V.V.: {\em Hankel operators in the theory of perturbations
of unitary and selfadjoint operators}, Funktsional. Anal. i
Prilozhen. 19 (1985), no. 2, 37--51.

\bibitem{Pot}
Potapov, V.P.:{\em The multiplicative structure of $J$-contractive matrix functions} (Russian),
Trudy Moskov. Mat. Obshch. 4 (1955), 125--236.

\bibitem{Ryb83}
Rybkin, A. V.:
{\em Trace formulas for resonances},
Teoret. Mat. Fiz.  56  (1983), no. 3, 439--447.

\bibitem{Ryb84}
Rybkin, A. V.:
{\em The spectral shift function for a dissipative and a selfadjoint 
operator, and trace formulas for resonances},
  Mat. Sb. (N.S.)  125(167)  (1984),  no. 3, 420--430.

\bibitem{Ryb85}
Rybkin, A. V.:
{\em  The discrete and the singular spectrum in the trace formula 
for a contractive and a unitary operator}, 
Funktsional. Anal. i Prilozhen.  23  (1989),  no. 3, 84--85

\bibitem{Wei1}
J. Weidmann, Lineare Operatoren in Hilbertr\"{a}umen. Teil II:
Anwendungen, 
B. G. Teubner, Stuttgart, 2003.

\bibitem{Y}
Yafaev, D.R.: {\em Mathematical Scattering Theory: General Theory},
Translations of Mathematical Monographs, Vol. 105, American Mathematical
Society, Providence, RI, 1992.

\end{thebibliography}
\end{document}